\documentclass[journal]{IEEEtran}

\usepackage{xcolor, soul}
\usepackage{graphicx} 
\usepackage{amsmath}
\usepackage{amssymb}
\usepackage{amsfonts}
\usepackage{subcaption}
\usepackage{algorithm}
\usepackage{algpseudocode}
\usepackage{booktabs}
\usepackage{url}
\usepackage{hyperref}
\usepackage{setspace}
\usepackage{cite}
\usepackage{verbatim}

\usepackage{pgfplots}
\pgfplotsset{compat=1.18}

\newcommand{\bc}[1]{\textcolor{blue}{#1}}
\renewcommand{\bc}[1]{#1}

\title{Disruption-aware Microservice Re-orchestration for Cost-efficient Multi-cloud Deployments}

\author{{Marco Zambianco}\thanks{M. Zambianco and S. Cretti are with Fondazione Bruno Kessler, Italy. D. Siracusa is with University of Trento, Italy. Corresponding author: M. Zambianco (email: mzambianco@fbk.eu)}, Silvio Cretti, Domenico Siracusa}

\IEEEpubid{\parbox{\textwidth}{\centering
    \vspace{-51cm} 
    \scriptsize This article has been accepted for publication in IEEE Transactions on Services Computing. This is the author's version which has not been fully edited and
content may change prior to final publication. Citation information: DOI \url{https://doi.org/10.1109/TSC.2025.3604373}
}}

\begin{document}

\maketitle



\begin{abstract}
Multi-cloud environments enable a cost-efficient scaling of cloud-native applications across geographically distributed virtual nodes with different pricing models. In this context, the resource fragmentation caused by frequent changes in the resource demands of deployed microservices, along with the allocation or termination of new and existing microservices, increases the deployment cost. Therefore, re-orchestrating deployed microservices on a cheaper configuration of multi-cloud nodes offers a practical solution to restore the cost efficiency of deployment. However, the rescheduling procedure causes frequent service interruptions due to the continuous termination and rebooting of the containerized microservices. Moreover, it may potentially interfere with and delay other deployment operations, compromising the stability of the running  applications. 
To address this issue, we formulate a multi-objective integer linear programming (ILP) problem that computes a microservice rescheduling solution capable of providing minimum deployment cost without significantly affecting the service continuity. At the same time, the proposed formulation also preserves the quality of service (QoS) requirements, including latency, expressed through microservice co-location constraints. Additionally, we present a heuristic algorithm to approximate the optimal solution, striking a balance between cost reduction and service disruption mitigation. We integrate the proposed approach as a custom plugin of the Kubernetes (K8s) scheduler. Results reveal that our approach significantly reduces multi-cloud deployment costs and service disruptions compared to the benchmark schemes, while ensuring QoS requirements are consistently met.

\end{abstract}

\begin{IEEEkeywords}
Microservice re-orchestration, cost minimization, service disruption minimization, multi-cloud systems, Kubernetes scheduler, optimization
\end{IEEEkeywords}

\section{Introduction}

In recent years, the multi-cloud computing paradigm has become increasingly adopted to distribute workloads across multiple public cloud regions that offer on-demand computational resources at different prices \cite{hong2019overview}. This business model gives service providers the opportunity to dynamically lease virtual nodes to accommodate the deployment of their cloud-native applications with limited operational overhead \cite{alonso2023understanding}.
In this scenario, optimizing resource utilization becomes crucial for reducing the economic cost of running containerized microservices on rented nodes across multiple cloud providers \cite{georgios2021exploring}. To facilitate this objective and, in general, the management of the microservices life-cycle, container orchestration tools such as Kubernetes (K8s) have become the standardized approach to automatize the allocation of workloads on the available nodes based on their resource requirements \cite{vayghan2018deploying} \cite{senjab2023survey}.

Leveraging these frameworks, the research activity has extended single-cloud orchestration solutions such as \cite{tamiru2021mck8s,wang2022container,monaco2023shared,alelyani2024optimizing}, which ignore the concept of deployment cost,  to achieve cost-efficient application deployments while ensuring that QoS requirements are met.
In particular, multi-cloud orchestration approaches primarily consist of scheduling algorithms that determine where to deploy microservices in order to reduce the deployment cost according to a fixed configuration of resource requirements and node availability within different geographically-distributed cloud regions \cite{rac_cost-aware_2024,luo_achieving_2024,shi_auto-scaling_2023}. At the same time, these schemes also ensure the co-location of delay-sensitive microservices within the same region to prevent QoS degradation caused by high inter-region network latency.

\begin{figure}[t]
    \centering
     \includegraphics[width=0.5\textwidth,trim={0 1.5cm 0 2cm},clip]{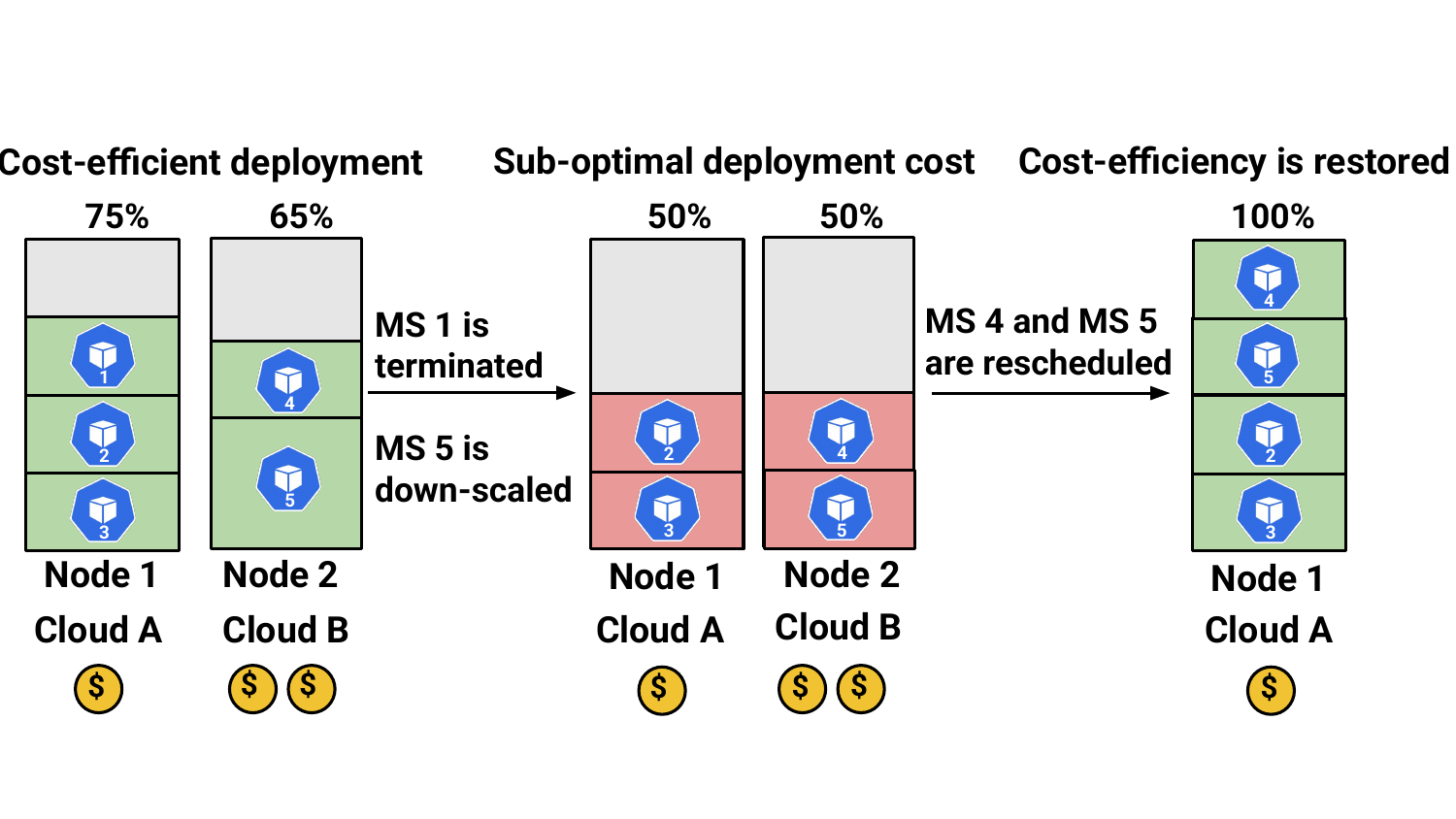}
     \caption{Example of microservice re-orchestration restoring the deployment cost efficiency.}
     \label{fig:res_frag_example}
\end{figure}

However, resource requirements of deployed microservices may change overtime due to various reasons including adjusting to more intensive task processing, managing seasonal spikes in user requests, accommodating the roll-out of new  microservice functionalities, mitigating the resource contention on overloaded nodes \cite{baarzi2021showar}. Similarly, resource availability on nodes can also fluctuate due to the deployment of new microservices to accommodate new applications or due to the termination of existing microservices that have completed their tasks, such as the training of machine learning models and processing of big data analytics \cite{xu2024practice}. The constant changes in resource requirements, along with the arrival and termination of microservices, progressively increase resource fragmentation, which in turn increases the demand of nodes needed to accommodate the re-configured applications \cite{larsson2023impact}. As a consequence, the deployment cost-efficiency steadily deteriorates, since the initial node assignment computed for the original configuration of microservices becomes gradually suboptimal in minimizing costs.

To address this issue, re-orchestrating microservices (in other words, the rescheduling of deployed microservices) offers a way to restore the cost-efficiency by re-allocating microservices on a cheaper configuration of multi-cloud nodes.  In Fig. \ref{fig:res_frag_example}, we provide a simple example to better illustrate this aspect. In detail, the cost efficiency of the microservice deployment deteriorates due to a combination of resource downscaling and workload termination (note that the same effect is produced by combining subsequent workload downscaling/upscaling with new workload arrival/termination). In this scenario, the unused resources, that were originally reserved, can now be utilized to re-orchestrate microservices from Node 2 to Node 1, thus reducing the deployment cost as Node 2 can be terminated.

An effective and practical microservice re-orchestration solution has to account for the impact on service continuity caused by the rescheduling process.
The rescheduling of deployed microservices is implemented by terminating existing containers and subsequently restarting them on the newly assigned destination nodes. This process causes service downtime which might not be tolerable for some applications since rescheduled microservices temporarily stop processing tasks. The severity of service disruption is exacerbated by the number of rescheduled microservices as the amount of pending unprocessed tasks increases. In addition, the re-orchestration of microservices also affects the system readiness to accommodate upcoming modifications to the configuration of deployed applications. In highly dynamic environments, the rescheduling procedure might conflict with the deployment operations (e.g., modification of environment variables, update of container images, interaction with monitoring and logging services, changing security policies) and may leave applications in an inconsistent and/or unresponsive state for an extended period of time \cite{joseph2021nature}\cite{jang2023enhancing}. For this reason, it is important to complete the rescheduling procedure in a short amount of time to minimize the impact on the deployment stability.
Overlooking these effects significantly reduces the benefit of potential cost savings achieved by the microservice re-orchestration, since frequent service interruptions undermine the dependability of the deployed applications. In this regard, existing multi-cloud orchestration schemes for cost minimization are designed under the assumption that microservices are scheduled for the first time, hence they are unsuited for microservice re-orchestration since they ignore the service disruption.

To fill this gap, we propose a disruption-aware re-orchestration algorithm that mitigates service disruption when rescheduling microservices. Our approach aims to minimize deployment costs in multi-cloud environments while preserving both service continuity and QoS requirements. Specifically, we design our rescheduling solution assuming that microservices are rescheduled using a rolling update strategy, which is a deployment option provided by the Kubernetes orchestrator. This mechanism reduces service downtime during rescheduling by first deploying an instance of rescheduled microservices on the new destination nodes before terminating the corresponding replicas at their original location \cite{singh2017container}. From a resource allocation perspective, this effect requires accounting for the coexistence of the new and old microservice instances during the rescheduling phase, temporarily increasing the node resource utilization, in order to achieve an optimal rescheduling decision. We previously leveraged this rescheduling strategy to re-orchestrate at runtime microservices in order to minimize the deployment cost as described in \cite{zambianco2024cost}. In this work, we extend our former re-orchestration solution to also include the minimization of the number of rescheduled microservices and of the rescheduling duration. These additions allow to further mitigate the service disruption severity by avoiding unnecessary microservice migrations and overextended duration of the rescheduling procedure. 
Moreover, unlike the simulated environment used to assess the results of our previous work, we enhance the performance evaluation by implementing the proposed re-orchestration solution as a custom Kubernetes scheduler plugin. This allows us to benchmark the results against the default Kubernetes scheduler configuration, providing a more accurate performance assessment for practical scenarios. In detail, we summarize our contributions as follows:
\begin{itemize}
    \item We formulate a multi-objective integer linear programming (ILP) problem, based on a customized version of the Variable Cost Bin Packing Problem, to compute the optimal rescheduling strategy that minimizes i) the total economic cost associated to multi-cloud nodes hosting microservices, ii) the number of rescheduled microservices and iii) the number of rescheduling slots. The first term addresses the cost minimization objective, where the second and third terms minimize service disruption. Moreover, we preserve the QoS performance by means of rescheduling constraints that enforce the co-location of communicating microservices having stringent latency requirements on nodes within the same cloud region. 
    \item We propose a low-complexity heuristic algorithm to approximate the optimal solution using an iterative greedy approach. In detail, the designed algorithm maximizes the number of unused high-cost nodes by parallelizing the rescheduling of microservices from those nodes to the cheapest available nodes without violating co-location requirements. This strategy achieves a balanced trade-off between cost reduction, number of rescheduled microservice and rescheduling duration.  
    \item We integrate the optimal and heuristic re-orchestration solutions as custom Kubernetes scheduler plugins in order to evaluate their performance on practical scenarios. Compared to the default Kubernetes scheduler implementation and an affinity-based cost-aware orchestration scheme, the proposed schemes can provide higher cost reductions using a lower number of rescheduled microservices and a shorter rescheduling duration while guaranteeing a consistent fulfillment of QoS requirements.
\end{itemize}

The remainder of the paper is organized as follows. Section II covers the related work. In Section III, we outline the considered system model. Section IV presents the optimal microservice re-orchestration solution and a heuristic algorithm for approximating it. In Section V, we analyze the obtained results. Finally, we draw the conclusion in Section VI.

\section{Related work}

The advent of multi-cloud computing has increased the complexity of orchestration schemes due to the distributed nature of such system. An overview of the main challenges and proposed solutions in the context of multi-cloud orchestration can be found in \cite{tomarchio2020cloud} and in \cite{wang2020survey}. Given the focus on cost minimization, we compare our work with respect to microservice orchestration solutions that propose scheduling strategies to reduce deployment costs while balancing resource utilization and service performance.

 The authors of \cite{aznavouridis2022micro} propose a graph clustering method for microservice placement in public clouds to optimize node resource usage and bandwidth, resulting in  reduction of  the associated deployment cost. Along similar lines, \cite{li2024carokrs} and \cite{ding2022kubernetes} propose Kubernetes-based scheduling strategies aimed at improving resource efficiency and reducing deployment costs. They achieve these goals 
 by respectively enhancing load balancing across nodes and by optimizing microservice placement through a genetic algorithm that considers shared library usage to minimize resource consumption.
While these works offer different strategies for minimizing deployment costs by essentially improving the resource utilization efficiency of microservices, they are static approaches. Microservices are scheduled by assuming a fixed configuration of resource requirements and resource availability on the nodes. As a result, the performance of such solutions in scenarios where the increasing resource fragmentation deteriorates the deployment cost efficiency remains uncovered. In contrast, we specifically focus the latter problem and we design an optimization-based rescheduling strategy to efficiently repack existing workloads on the cheapest multi-cloud node configurations based on the current resource availability.


In the context of dynamic resource re-configuration, \cite{jiang2020cloud} and \cite{sfakianakis2021skynet} propose orchestration frameworks that enhance resource utilization for cloud-native, data-intensive applications by minimizing resource fragmentation across geo-distributed nodes and by employing control-theoretic adaptive controllers to continuously right-size microservice resources while preserving QoS, respectively. 
 A similar strategy is proposed in \cite{wang2020elastic}, which combines microservice scaling and allocation to optimally minimize the number of required virtual nodes based on the current resource consumption of each application. With focus on microservice rescheduling,  \cite{aldwyan2021elastic} proposes a microservice scheduling framework, based on a genetic algorithm, that relocates microservice on multi-cloud clusters in order to jointly minimize the deployment cost and service latency. Alternatively, \cite{khan2022automatic} introduces a dynamic resource management scheme that migrates deployed workloads based on their job execution time using an unsupervised learning approach. Their solution improve data center utilization, reducing the deployment cost by lowering energy expenditure. In addition to the tradeoff between cost and latency, the authors of \cite{bracke2024multiobjective} accounts for resource overloading and propose a multi-objective rescheduling strategy to improve application service time by co-locating interdependent microservices on the same node while also minimizing resource contention. 
 The aforementioned solutions address the limitations of static orchestration by rescheduling microservices to improve resource utilization, taking into account factors such as resource balance, QoS requirements, and costs. However, they lack strategies to mitigate service interruptions caused by frequent microservice re-allocations. In contrast, our approach minimizes service disruptions during rescheduling by combining the usage of a rolling update deployment strategy with the minimization of the number of rescheduled microservices and rescheduling duration, ensuring cost-efficient deployments with minimal impact on service continuity.


Regarding orchestration solution focusing primarily on QoS fulfillment over costs, the authors of \cite{zhong2020cost} and \cite{sampaio2019improving}  propose heuristic-based microservice rescheduling schemes to automatically co-locate microservices having high affinity values in terms of number of exchanged data packets in order to further improve the cost mitigation. Similarly, a Kubernetes scheduler plugin is developed in \cite{zhang2021zeus} that adaptively balances the resource allocation  of best-effort and latency-sensitive workloads. The aforementioned schemes enforce co-location requirements by imposing affinity rules between microservices, acting as soft constraints which might not be satisfied at runtime. In the context of microservice rescheduling, this approach performs poorly, since the co-locations requirements satisfied by the current deployment configuration could be possibly violated once the rescheduling phase is completed. For this reason, we first design an optimization problem where co-location requirements work as hard constraints to limit the possible rescheduling options, ensuring an optimal rescheduling configuration that reduce costs and guarantee QoS performance. This strategy is also embedded within the heuristic algorithm that approximates the optimal solution to prioritize QoS fulfillment.

Finally, in our previous work \cite{zambianco2024cost}, we proposed a re-orchestration algorithm that optimizes microservice rescheduling via a rolling-update strategy to minimize deployment costs while mitigating service disruption and preserving co-location requirements. However, its assumption of a fixed number of rescheduled microservices and sequential rescheduling introduced inefficiencies, leading to excessive migrations and prolonged deployment times in large-scale systems. To overcome these limitations, we jointly minimize the service disruption severity and the deployment cost to offer a more balanced rescheduling strategy.

\section{System model}
We introduce the main elements of the considered model that is composed of the multi-cloud architecture and by the microservice re-orchestrator module.

\subsection{Multi-cloud microservice deployment model}

We consider $R$ geographically-distributed node clusters, each one owned and administrated by a different cloud provider (e.g., AWS, Microsoft Azure, Google Cloud) that offers computational resources in the form of virtual nodes. We indicate the set of virtual nodes hosted by each individual multi-cloud cluster as $N_0,...,N_{R-1}$ whereas we represent the set of all virtual nodes as $N=\bigcup_{i=0}^{R-1} N_i$. Each node provides a maximum of $C_n = \{c^{(cpu)}_n, c^{(ram)}_n \}$ computational resources expressed as CPU resources and RAM resources, respectively.  Due to the geo-distribution of the cloud infrastructure, virtual nodes belonging to different clusters experience some non-negligible latency denoted as $D_{n,n'} > 0$, $n \in N_i$ , $n' \in N_{j \neq i}$. Conversely, we assume that intra-cluster network latency can be approximated as zero, thus $D_{n,n'} = 0$, $n, n' \in N_i$.
The various nodes can be rented by some service provider at fixed price of $p_n C_n $, with $p_n$ indicating the price per unit of resources, to deploy microservice-based applications encompassing different use cases such as e-commerce platforms, content streaming services, machine learning tasks. We remark that the price associated to each node is associated with the leasing of the whole instance, hence it is constant regardless of the amount of computational resource utilized by allocated microservices (i.e., 10\% utilization rate of the node resources has the same cost of 90\% utilization rate). We indicate the set of deployed applications as $A$ and the set of microservices composing each application as $M_a, a \in A$. Moreover, we define the set of all microservices as $M = \bigcup_{a \in A} M_a$. Each microservice requires a minimum of $r_{a,m}=\{ r^{(cpu)}_{a,m},r^{(ram)}_{a,m} \}$ resources in order to be scheduled on a given node. We represent the assignment of microservices on the various virtual nodes using the indicator variable $s_{a,m,n} = 1$ if $m \in M_a, a \in A$ is allocated on node $n \in N$, $s_{a,m,n} = 0$ otherwise.

\begin{figure}[t]
    \centering
     \includegraphics[width=0.5\textwidth]{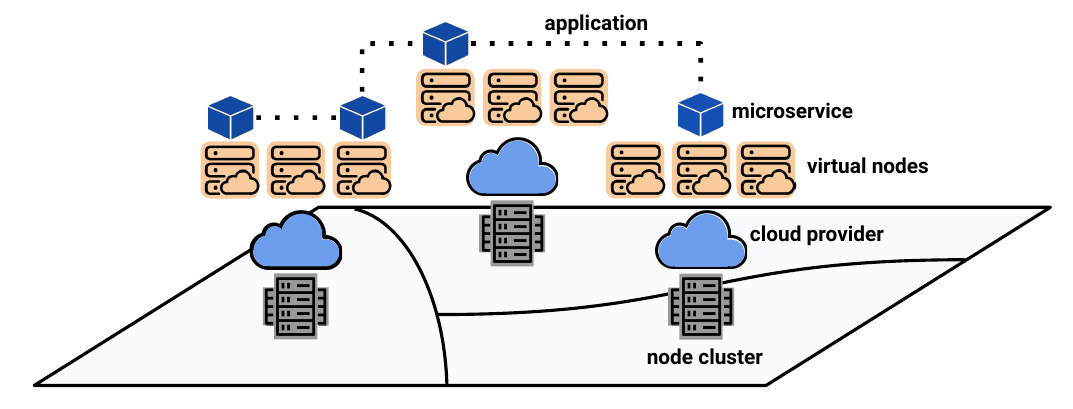}
     \caption{Multi-cloud system model. Microservices of each application are deployed on various virtual nodes belonging to different geographically-distributed cloud providers.}
     \label{fig:sys_model}
\end{figure}

Moreover, to prevent the degradation of QoS performance caused by high inter-region network latency, microservices may be subject to some locality constraints which enforce their co-location on virtual nodes within the same cloud region. In detail, we refer to as $d_{a,m,m'}$ the maximum tolerable network latency between communicating microservices $m,m' \in M_a$ of application $a \in A$. Then, we express the location feasibility for each pair of communicating microservices by defining the indicator variable $\ell^{(n,n')}_{a,m,m'} = 1$ if $d_{a,m,m'} \geq D_{n,n'} \quad  \forall m,m' \in M_a, \forall n, n' \in N$, $\ell^{(n,n')}_{a,m,m'} = 0$ otherwise. In other words, $\ell^{(n,n')}_{a,m,m'} = 1$ indicates that the additional network latency $D_{n,n'}$ that affects microservices $m$ and $m'$ on virtual nodes $n$ and $n'$ is below the maximum threshold $d_{a,m,m'}$ and thus the allocation of related microservices satisfies the QoS requirements.
Based on this notation, we analytically represent a feasible deployment configuration as 
\begin{multline}
S_A = \{ s_{a,m,n} : \sum_{a' \in A} \sum_{m' \in M_a} s_{a',m',n'} r_{a',m'} \leq C_n \quad \land  \\
    s_{a,m',n'} = 1 \quad \text{if} \quad  \ell^{(n,n')}_{a,m,m'} = 1, \forall m' \in M_a, n' \in N \}  
\end{multline}
Essentially, from the above definition, a feasible microservice deployment must not exceed the maximum capacity of the virtual nodes and, at the same time, must satisfy the co-location constraints.

\subsection{Disruption-aware microservice re-orchestration  model}

\begin{figure*}[t]
    \centering
     \includegraphics[width=0.9\textwidth,trim={0 3cm 0 0},clip]{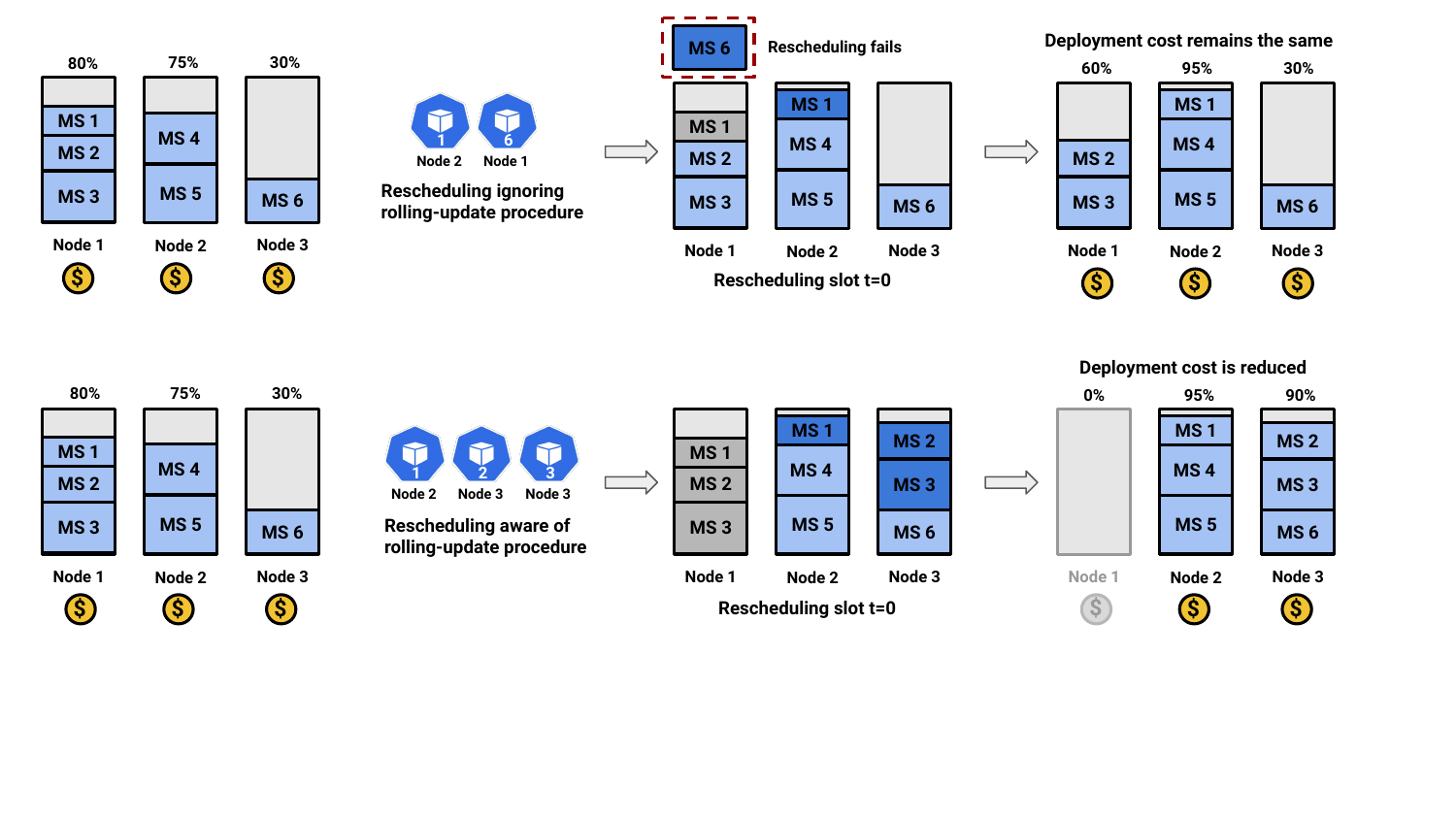}
     \caption{Example showing the effect of rolling-update deployment strategy on deployment cost. Microservices in dark blue are the new instances being rebooted on the designated destination nodes. Microservices in gray are the old instances that are going to be terminated once the rebooting process of the new instances is completed.} 
     \label{fig:resch_example}
\end{figure*}

We consider a multi-cloud environment where the resource requirements and/or the number of microservices deployed change over time. The service provider continuously monitors the cost-efficiency of the deployment and triggers a microservice re-orchestration at any point during the deployment life-cycle to restore cost optimality. 
Examples of re-orchestration trigger conditions include low resource utilization on some nodes, deployment cost exceeding a maximum cost budget, and availability of less expensive virtual nodes in a different cloud region.
The re-orchestrator module is responsible for computing a new feasible deployment configuration that reallocates microservices onto a more cost-effective set of active virtual nodes.  This rescheduling strategy ensures that cost is bounded and cannot increase during the microservice migration process as only a subset of the already available nodes are eventually selected as destination nodes for the new deployment configuration. Nodes left empty after the rescheduling phase are terminated, thus reducing overall costs.
We model the execution of the rescheduling phase as a sequence of time slots in which one or multiple microservices are reallocated on the new destination nodes. In particular, indicating with $T$ the set of time slots, the maximum number of time slots available is $|T| = |\bigcup_{a \in A} M_a|$,  which corresponds to the limit scenario where all deployed microservices are individually rescheduled in a different time slot.
The rescheduling phase affects the application service continuity due to the inability to process tasks during the time period in which the container of a rescheduled microservice is restarted on the new destination node. \bc{The severity of such disruptions is exacerbated by the number of functional dependencies among microservices (i.e., dependencies in which a microservice requires the output of one or more other microservices to operate correctly). Consequently, the rescheduling of even a small number of microservices responsible for providing processed tasks may lead to significant degradation in overall service performance.}
For example, time-sensitive applications may fail to process some tasks within the time deadline or machine learning pipelines may lose some training progress due to unresponsive microservices. 
Similarly, the duration of the rescheduling process can also affect applications' stability and interfere with the deployment operations, causing delays and further service interruptions. 
Although temporarily, this effect hinders the application performance as a part of the deployed microservices is not fully operational. 
Consequently, in addition to the final goal of restoring the cost-efficiency of deployment
while preserving the co-location constraints, the computed microservice re-orchestration is also disruption-aware as it mitigates the severity of service disruption caused to the applications by the rescheduling phase. In detail, we assume that the minimization of service disruption can be achieved by a rescheduling of microservices that satisfies three objectives:
\begin{itemize}
    \item \textbf{Rolling-update  redeployment strategy}: microservice rescheduling is operationally executed by first redeploying each microservice instance on the new node and, only once the rebooting process is completed,  the old instance is terminated. \bc{We remark that the extra resource usage during rolling updates is safely managed by modern orchestrator tools such as K8s to minimize the strain on both the control plane and the underlying network infrastructure. For example, K8s limits a large number of simultaneous rolling updates to smooth out memory and CPU spikes. Similarly, it gradually shifts network traffic by steering packets to new microservices only when they are fully ready. These mechanisms ensure minimal control plane overload and network congestion while delivering close-to-zero downtime and seamless microservice re-deployment.}

    \item \textbf{Minimization of the number of rescheduled microservices}: the number of rescheduled microservices is minimized to prevent unnecessary migrations that do not have benefits in terms of cost reduction, but can still affect service continuity due to delays in tasks execution.  
    \item \textbf{Minimization of rescheduling phase duration}: the number of rescheduling slots is minimized to reduce possible conflicts with the deployment operations, possibly affecting the stability of applications and impairing their functionalities.
\end{itemize}

The impact of service disruption varies based on each microservice's role and redundancy. Front-end services or those without replicas are more affected by rescheduling, whereas replicated services experience minimal disruption. Considering these factors adds complexity to re-orchestration and is left for future analysis.

\subsection{Impact of rolling-update deployment strategy on cost}

The combination of a rolling-up update deployment strategy with minimization of the number of rescheduled microservices and rescheduling slots requires proper modeling of their dynamics to optimally minimize the deployment cost.
During the rolling-update process, both old and rescheduled microservice instances are temporarily active on the original and destination nodes, respectively. The higher resource usage on the nodes increases the resource contention, that hampers the effectiveness of cost minimization by making the selection of which microservices to reschedule less obvious. In Fig. \ref{fig:resch_example}, we illustrate an example of microservice rescheduling to show how ignoring the rolling-update procedure can impact the deployment cost. The top picture shows a  re-orchestration scheme that ignores the rolling-update rescheduling dynamic, hence it assumes that microservices can be instantaneously rescheduled on the destination nodes. Given the initial deployment configuration, the rescheduling of Microservice 1 on Node 2 and Microservice 6 on Node 1 would lower the deployment cost and would provide minimum service disruption as two microservices are rescheduled within the same rescheduling slot. However, such a solution fails in practice, since the old instance of Microservice 1 is still active and prevents Microservice 6 from being rescheduled on Node 1. As a result, Microservice 6 remains allocated on Node 3, resulting in the same deployment cost as in the initial deployment configuration. Differently, the bottom picture shows a  re-orchestration scheme that accounts for the rolling-update rescheduling dynamic when computing the microservice rescheduling. In this scenario, microservice 1, 2, and 3 are now rescheduled on Node 2 and Node 3. This solution overcomes the resource contention caused by the old microservice instances and successfully reduces the deployment cost through the termination of Node 1. 

Given the impact of the rolling-update mechanism on cost performance, it is important to model this dynamic to design an effective re-orchestration scheme that optimally minimizes the deployment cost. To address this, the next section introduces an optimization problem that incorporates the effects of the rolling-update process on node resource utilization, ensuring that the proposed microservice rescheduling can be executed in practice.

\section{Problem formulation}
We present the multi-objective optimization problem formulation that computes the optimal microservice rescheduling solution, minimizing costs while mitigating service disruption and preserving QoS requirements. Moreover, we also propose a heuristic algorithm to approximate the optimal formulation.

\subsection{Optimal re-orchestration solution}

We model the cost minimization problem as a variation of Variable Cost Bin Packing problem where we extend the original formulation to include the minimization of service disruption and the fulfillment of QoS. We analytically model the rescheduling of microservices on the available virtual nodes introducing the binary variable $x^{t}_{a,m,n}  = 1$, which indicates that microservice $m$ of application $a$ is rescheduled on node $n$ at time slot $t$,  $x^{t}_{a,m,n}  = 0$ otherwise. Similarly, we introduce three optimization variables $y_n, v_{a,m}$ and $z_t$ to evaluate the deployment cost, the number of rescheduled microservices, and the number of rescheduling slots, respectively.  
In detail, the binary variable $y_n =1$ indicates that virtual node $n \in N$ is hosting any of the deployed microservices, whereas $y_n = 0$ indicates that node resources are not used by any microservice and therefore it can be terminated to reduce the deployment cost.  
The binary variable $v_{a,m} = 1$ indicates that microservice $m \in M_a$ of application $a \in A$ is rescheduled on a different virtual node, whereas $v_{a,m} = 0$ otherwise. Note that $v_{a,m}$ can be considered as a simplified version of the variable $x^{t}_{a,m,n}$ ignoring the destination node and time slot information and it is used to streamline the formulation of the objective function. Finally, the binary variable $z_t = 1$ indicates that time slot $t$ is dedicated to reschedule at least one of the deployed microservices. 
Using these variables, we formulate the following optimization problem. 

\begin{equation} \label{eq_obj_func}
   \min_{v,y,z} \sum_{n \in N} p_n C_n \cdot y_n +  \alpha \Big( \sum_{a \in A}\sum_{m \in M_a} v_{a,m} + \sum_{t \in T} z_t \Big)
\end{equation}

Subject to

\begin{equation}\label{eq_cons1}
    \sum_{a \in A} \sum_{m \in M_a} \sum_{t \in T} r_{a,m} x^{t}_{a,m,n} \leq y_n \Delta C_{n}  \quad \forall n \in N
\end{equation}

\begin{equation}\label{eq_cons2}
    \begin{split} \sum_{t \in T} x^{t}_{a,m,n}  + \sum_{t \in T} x^{t}_{a,m',n'} \leq 1 + \ell^{(n,n')}_{a,m,m'} \\ \forall m,m' \in M_a,  \forall a \in A, \forall n,n' \in N \end{split}
\end{equation}

\begin{equation}\label{eq_cons3}
\begin{split}
     \sum_{t' \in T_{t'< t}} \Big{\{} \sum_{n' \in N} \Big[ \sum_{a \in A} \sum_{ m \in M_a}   s_{a,m,n'} \Big( x^{t'}_{a,m,n}   r_{a,m} \Big) - \\ \sum_{a' \in A} \sum_{m' \neq m \in M_{a'}} 
                                         s_{a',m',n} \Big( x^{t'}_{a',m',n'}   r_{a',m'}  \Big)  \Big] \Big{\}} \leq \Delta C_n
                                      \\ \quad \forall n \in N, \forall t \in T
\end{split}
\end{equation}

\begin{equation}\label{eq_cons8}
    v_{a,m} = s_{a,m,n} - \sum_{t \in T} x^{t}_{a,m,n} \quad \forall a \in A, \forall m \in M, \forall n \in N 
\end{equation}

\begin{equation}\label{eq_cons9}
    z_t \leq \sum_{m \in M_a}\sum_{n \in N} x^{t}_{a,m,n}  s_{a,m,n}  \leq  z_t \cdot T \quad  \forall a \in A, \forall t \in T  
\end{equation}

\begin{equation}\label{eq_cons4}
   \sum_{t \in T} \sum_{n \in N}  x^{t}_{a,m,n} = 1 \quad \forall m \in M_a, \forall a \in A
\end{equation}

\begin{equation}\label{eq_cons5}
   z_{t} \geq z_{t+1} \quad \forall t \in T
\end{equation}

\begin{equation}\label{eq_cons7}
\begin{split}
       y_n, x^{t}_{a,m,n}, v_{a,m}, z_{t} \in \{0,1\} \quad  \forall a \in A, \forall m \in M_a,  \\ \forall n \in N, \forall t \in T
\end{split}
\end{equation}

The objective function \eqref{eq_obj_func} includes three components. The first addend computes the total deployment cost as a summation of the costs associated to each active multi-cloud virtual node. The second and third addends quantify the level of service disruption produced by the rescheduling process in the form of a penalty (note that QoS fulfillment is instead enforced by the co-location constraints)  that depends on the total number of rescheduled microservices and the total number of needed rescheduling slots, respectively. The impact of these values can be modulated by tuning the parameter $\alpha \in [0,1]$, that we refer to as disruption cost. Intuitively, higher values of $\alpha$ promote a microservice rescheduling solution that prioritizes service continuity over cost minimization.

Problem constraints, beside ensuring the solution feasibility in terms of resource occupation, implement the rolling update deployment strategy and enforce colocation requirements. In detail, the constraint \eqref{eq_cons1} ensures that the new resource occupation of the deployment does not exceed the assigned node capacity, where $\Delta C_n = C_n - \sum_{a \in A}\sum_{m \in M} s_{a,m,n} r_{a,m}, \forall n \in N$ corresponds to the available capacity on each virtual node according to the initial deployment configuration $S_A$. Moreover, it links the cost optimization variable $y_n$ with the rescheduling optimization variable $x^{t}_{a,m,n}$ by ensuring that empty nodes cannot accommodate microservices. Constraint \eqref{eq_cons2} enforces colocation constraints by guaranteeing that a given microservice pair $m,m'$ cannot be allocated on nodes $n$ and $n'$ in different cloud regions if $\ell^{(n,n')}_{a,m,m'} = 0$. Constraint \eqref{eq_cons3} accounts for the coexistence of the old and new microservice instance $m$ rescheduled on a different node $n' \neq n$ on time slot $t$ due to the rolling-update deployment strategy. This is achieved in practice by ensuring that the variation in resource occupation in every node $n$ never exceeds the related residual capacity $\Delta C_n$ in each rescheduling slot $0,..,t',..t$. Specifically, the first term inside the square brackets computes the total amount of resources consumed by microservices that were rescheduled on node $n$ during any of the previous rescheduling slots $0, ..., t'$. In contrast, the second term computes the total amount of resources freed up on node $n$ as a result of microservices being rescheduled to other nodes during time slots  $0, ..., t'$. Note that the resources of microservices rescheduled on the same node (i.e., when $n = n'$) are not included, as they are in fact not rescheduled in practice. Constraint \eqref{eq_cons8} ensures that a microservice is considered as rescheduled only if the destination node indicated by $x^{t}_{a,m,n}$ differs from the original node assignment indicated by the initial deployment configuration $s_{a,m,n} = 1$.
Constraint \eqref{eq_cons9} bounds the number of rescheduling slots up to $T$ which is the total number of microservices currently deployed and guarantees that multiple microservices can be rescheduled within the same time slot $t$.
Constraints \eqref{eq_cons4} and \eqref{eq_cons5}  guarantee that every microservice is uniquely assigned to a single node only and that the number of employed rescheduling slots are contiguous, respectively. 
Finally, constraint \eqref{eq_cons7} expresses the integer nature of the problem.

Intuitively, the computational complexity of the proposed problem is NP-Hard as it can be reduced in polynomial time to the formulation of the classical Bin Packing problem \cite{martello1990lower}. The latter represents the corner scenario where all microservices are rescheduled at the same time to minimize the number of equally-priced virtual nodes belonging to a single cloud region. Therefore, due to its poor scaling capabilities, the resolution time is prohibitive for large-scale deployment scenarios composed by a high amount of microservices deployed across many multi-cloud virtual nodes. For this reason, in the next section, we lower the computation complexity by designing a heuristic approach for re-orchestrating microservices.

\subsection{Heuristic re-orchestration algorithm}

The proposed heuristic algorithm strikes a balanced trade-off between the various optimization objectives. This approach eliminates the need for fine-tuning additional parameters that would otherwise depend on the specific microservice deployment configuration and infrastructure, making it easier to apply effectively in real-world scenarios. The general idea is to decompose the computation of the various objectives (cost, number of rescheduled microservices, number of rescheduling slots) while still ensuring that their minimization does not conflict with one another and that co-location requirements are consistently met. To reach this goal, as highlighted in the previous example in Fig. \ref{fig:resch_example}, it is important to mitigate the resource contention during the rescheduling process caused by the rolling-update mechanism as it could deny the rescheduling of some microservices, preventing the achievement of a more cost-efficient node configuration.

\begin{algorithm}[t]
\caption{Heuristic}\label{alg:heuristic}
\begin{algorithmic}[1]
    \State \textbf{Input:} $s_{a,m,n},p_n,C_n,\Delta C_n, \ell^{(n,n')}_{a,m,m'}, {r}_{a,m}$
    \State \textbf{Output:} $x^{t}_{a,m,n}$ 
    \State Initialize $x^{t}_{a,m,n} \!=\! 0 \quad  \forall m \!\in\! M_a, \forall a \!\in\! A, \forall n \!\in\! N, \forall t \!\in\! T$ \label{algo:init_begin}
    \State Initialize $y_{n} = 1 \quad \forall n \in N$ 
    \State Initialize rescheduling time slot $t=0$ 
    \State Initialize $\Delta C^{\text{roll}}_{n} = \Delta C_{n}$
    \State Initialize $P_{\text{prev}} = +\infty$  and $P_{\text{current}} =  \sum_{n \in N} p_n C_n$ \label{algo:init_end}
    \While{$P_{\text{current}} < P_{\text{prev}}$} \label{algo:main_algo_begin}
        \State Initialize $L_{N}(n) = [0, ..., n, ... , N-1: n \in N]$ \label{algo:sort_begin}
        \State Sort $L_{N}$ in ascending order of price $p_nC_n$
        \For{\textbf{each} $r \in R$}
            \State  Sort $L_{N_r}(n)$ in descending order of $\Delta C_n$
        \EndFor
        \State Initialize $\bar{L}_{N}(n) = [N-1, ..., n, ... , 0: n \in N]$ \label{algo:sort_end}
        \While{$L_{N}(n)$ is not empty} 
            \State Set $n_{min}$ = $L_{N}(0)$ \label{algo:select_min_node_begin}
            \State Remove $n_{min}$ from $L_{N}$ and  $\bar{L}_{N}$ \label{algo:select_min_node_end}    
            \For{\textbf{each} $\bar{n} \in \bar{L}_{N}(n)$}  
                \State Set $n_{max}$ = $\bar{L}_{N}(\bar{n})$ 
                \For{\textbf{each} $m \in M$ allocated on $n_{max}$} \label{algo:fill_min_node_begin}
                    \State{Identify $a \in A$ such that $m \in M_a$}
                   \If{$\Delta C^{\text{roll}}_{n_{min}} \!  \geq \! r_{a,m}$}
                   \If{$\ell^{(n_{min},n_{max})}_{a,m,m'} \! = \! 1, \quad  \forall  m' \! \in \! M_a$} 
                        \State Update $\Delta C^{\text{roll}}_{n_{min}} \leftarrow \Delta  C^{\text{roll}}_{n_{min}} - r_{a,m}$ 
                        \State Set $x_{a,m,n_{min}} = 1$  \label{algo:update_x_var_end}   
                    \Else{ Try reschedule  $\forall m' \in M_a$ \par \hspace{2.5cm} on $n \hspace{0.5em} \text{such that} \hspace{0.5em} \ell^{(n_{min},n)}_{a,m,m'} \! = \! 1$} \label{algo:try_schedule_coloc} 
                    \EndIf
                    \EndIf
                \EndFor
            \EndFor
        \EndWhile
        \For{\textbf{each} $n \in  N$}  
            \If{ $m \in M$ allocated on $n$ is rescheduled} \label{algo:update_delta_update_begin} 
                \State{Identify $a \in A$ such that $m \in M_a$}
                \State Update $\Delta C^{\text{roll}}_{n} \leftarrow \Delta  C^{\text{roll}}_{n} + r_{a,m}$
            \EndIf \label{algo:update_delta_update_end} 
             \If{$\Delta C^{\text{roll}}_{n} = C_n$} \label{algo:turn_off_max_node_begin}
                            \State Set $y_{n} = 0$
                            \State Set $P_{\text{current}} \leftarrow P_{\text{current}} - p_n C_n$ \label{algo:update_cost}
        \EndIf \label{algo:turn_off_max_node_end}
        \EndFor
        \State Update $t \leftarrow t + 1$ \label{algo:update_time_slot}
    \EndWhile \label{algo:main_algo_end}
\end{algorithmic}
\end{algorithm}

Following this observation, we adopt a greedy iterative approach that in each rescheduling slot exploits the resource available on the various nodes to progressively repack microservices from the highest cost node to the cheapest node that satisfies the co-location requirement (if any). In detail, the proposed strategy ensures a consistent and continuous QoS compliance since it prevents QoS violations by either restricting the migration of microservices with co-location requirements to the same cloud region or ensuring that they are simultaneously rescheduled to the new target region.
Moreover, the selection of the cheapest available nodes as destination node in combination with the strict utilization of the available resources in each slot also limits the number of rescheduled microservices, as the latter are selected with the goal of improving the resource utilization of the least expensive nodes. As a result, microservice are rescheduled only on nodes that can actually accommodate the required resources in a given slot without the need to preemptively rescheduling other microservices. 
We provide the pseudocode of the designed heuristic scheme in Algorithm \ref{alg:heuristic}. In detail, the algorithm takes as input the various parameters defining the deployment configuration and the multi-cloud infrastructure and provides as output the rescheduling of microservices on the available nodes. Lines \ref{algo:init_begin}-\ref{algo:init_end} initialize the various parameters needed for the computation: $\Delta C^{\text{roll}}_{n}$ tracks the resource availability in each node accounting for the rolling update deployment logic, $P_{\text{prev}}$ and $P_{\text{current}}$ corresponds to the deployment cost before and after the rescheduling procedure in the time slot $t$, respectively. Lines \ref{algo:sort_begin}-\ref{algo:sort_end} compute the list of nodes $L_{N}(n)$ sorted by increasing price, where equally priced nodes (i.e., nodes within the same cloud region) are sorted in decreasing order of resource occupation. Lines \ref{algo:select_min_node_begin}-\ref{algo:select_min_node_end} select the cheapest and most loaded node, denoted as $n_{min}$, which is the element in the first position in $L_{N}(n)$, and remove it from the list. Lines \ref{algo:fill_min_node_begin}-\ref{algo:update_x_var_end} progressively reschedule microservices allocated on the most expensive and least loaded node, denoted as $n_{max}$ and corresponding to the element in the last position in $L_{N}(n)$,  to node $n_{min}$ within the same time slot $t$. In particular, the rescheduling of microservice $m$ occurs only if there are enough resources available (both for CPU and RAM resources) at the destination node $n_{min}$, where the resource occupation $\Delta C^{\text{roll}}_{n}$ is updated taking into account the coexistence of old and new microservice instances. 
Moreover, to guarantee the preservation of co-location requirements, in line \ref{algo:try_schedule_coloc} we reschedule microservice $m$ only if co-located microservices can also be rescheduled on nodes within the same region of node $n_{min}$. Otherwise, the rescheduling is restricted to nodes belonging to the region of node $n_{max}$. 
Once $L_{N}(n)$ is empty, it means that there are no resource available in time slot $t$ to reschedule more microservices because of the rolling update procedure saturating the resources. Consequently, lines \ref{algo:update_delta_update_begin}-\ref{algo:update_cost} release the resources occupied by the old microservice instances that have been rescheduled and update the deployment cost $P_{current}$ by  subtracting the cost of the resulting empty nodes. Finally, the whole procedure \ref{algo:main_algo_begin}-\ref{algo:main_algo_end} is repeated as long as the rescheduling of the microservice in each time slot provides a reduction in deployment cost, otherwise it ends.
Intuitively, the proposed scheme enforces a coordinated microservice rescheduling to improve the resource utilization efficiency on the cheapest nodes by progressively migrating microservices originally allocated on most expensive nodes.  At the same time,  to minimize the number of rescheduled microservices, it prioritizes the re-orchestration of microservices from nodes with moderate resource consumption to nodes with higher resource consumption.

The computational complexity of the proposed approach can be computed as follows. The sorting of nodes based on price and resource occupation has complexity $O(N\cdot\text{log}N)$. The selection of microservices to reschedule has complexity $O(M)$ as it is performed by evaluating microservices as they are assigned to the ordered list of nodes (e.g., microservices allocated to the most expensive nodes that have the highest amount of resources available are checked first). Similarly, resource occupation on the various nodes is computed by iterating over nodes and removing the resources of the rescheduled microservices, thus it has complexity $O(N)$. Overall, one iteration of the algorithm has complexity $O(M \cdot N\cdot\text{log}N) + O(N) $, which can be further simplified asymptotically to $O(M \cdot N\cdot\text{log}N)$.

\section{Performance evaluation}

\subsection{Simulation setup}

We evaluated the performance of the \textbf{Optimal} and \textbf{Heuristic} solutions against two benchmark schemes:
\begin{itemize}
    \item \textbf{Kubernetes (K8s)}: this scheme progressively reschedules microservices belonging to the most expensive cloud regions using the default scheduling algorithm implemented in Kubernetes  orchestrator. The latter has become the standardized tool to manage containerized applications and thus offers a reliable baseline. We configured the Kubernetes scheduler by combining \textit{MostAllocated}, \textit{NodeAffinity} and \textit{PodAffinity} options together with a \textit{RollingUpdate} (RU) deployment strategy to achieve cost minimization, QoS requirements fulfillment and service continuity, respectively. A more detailed explanation of the considered options can be found in \cite{kube2023sched}. 
    \item \textbf{Bisecting K-Means} \cite{aznavouridis2022micro}: this scheme minimizes the deployment cost of microservices hosted in public cloud environments using a hierarchical clustering process that prioritizes the allocation of microservices with co-location requirements to balance costs and inter-microservice latency. However, in the original approach, microservices are recreated (in other words, they are terminated before being rescheduled on the destination nodes) resulting in service disruption. To better highlight the advantages of our disruption-aware approach, we also implement a variation of the scheme that uses a rolling-update deployment strategy to reduce service interruptions. We refer to the original version as \textit{Bisecting K-Means with Recreate} (B-RC), and the variation as \textit{Bisecting K-Means with RollingUpdate} (B-RU).
\end{itemize}

 \begin{figure}[t]
	\begin{center}
		\includegraphics[trim={0 0.8cm 0 0.8cm},clip,width=0.95\linewidth]{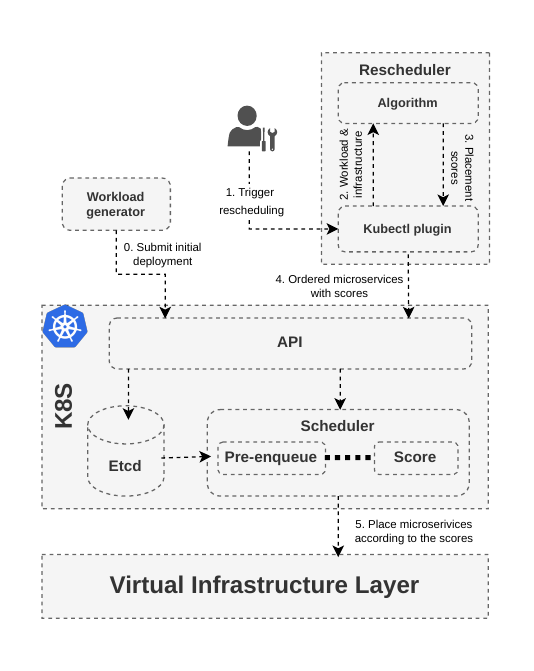}
		\caption{Software architecture of the emulation environment}
		\label{fig:architecture}
	\end{center}
\end{figure}

 We quantitatively evaluate the performance of our approach with respect to the benchmark schemes by computing the following metrics:
\begin{itemize}
    \item \textbf{Deployment cost}: this metric measures the cost efficiency of the new microservice configuration on the available nodes after the rescheduling phase and is represented in the form of cost reduction with respect to the initial deployment cost.
    \item \textbf{Rescheduled microservices}: this metric contributes to the evaluation of service disruption and measures the number of rescheduled microservices required to achieve the new deployment configuration.
    \item \textbf{Rescheduling duration}: this metric contributes to the evaluation of service disruption and measures the time required to execute the rescheduling of microservices (step 5 in Fig. \ref{fig:architecture}).
    \item \textbf{QoS fulfillment rate}: this metric assesses the effectiveness of the new deployment configuration in meeting QoS requirements by computing the fraction of satisfied co-location constraints after the microservice rescheduling.
\end{itemize}

We implemented the proposed Optimal and Heuristic schemes as plugins of the K8s scheduler. We also developed a custom emulation environment leveraging K8s orchestrator to manage the microservice rescheduling. The testbed was developed using a combination of Go and Python programming languages. In particular, we employed PySCIPOpt library to solve \eqref{eq_obj_func}-\eqref{eq_cons9}  using SCIP optimizer, which is a non-commercial solver for mixed ILP programming \cite{maher2016pyscipopt}. The code of the Kubernetes plugin and the emulation environment is publicly available at \url{https://gitlab.fbk.eu/fogatlas-k8s/kubectl-reschedule}.

To emulate various configurations of virtual nodes, microservices and resource requirements,
we employed the software toolkit Kubernetes WithOut Kubelet (KWOK) \cite{kwok2024}. This tool allows setting up fake virtual nodes, behaving like real ones, as well as to fully simulate the life-cycle of fake pods (pods are the atomic unit that can be managed by Kubernetes and can be considered as a microservice from a logical perspective). The advantage of this architectural design choice ensures a very low memory footprint that can be handled by a single physical machine. Nonetheless, since original control plane functionalities of Kubernetes, including the scheduler implementation, are preserved, we can reliably analyze the performance of our re-orchestration scheme without compromising the accuracy of results, ensuring their applicability in real-world scenarios.

\pgfplotsset{every axis/.append style={line width=1.5pt,tick style={line width=0.6pt}}}
\pgfplotsset{every axis/.append style={font=\small}}
\pgfplotsset{every axis y label/.append style={yshift=-4pt}}
\pgfplotmarksize=2pt
\begin{figure*}[t]
    \centering
    \begin{subfigure}[b]{0.25\textwidth}
        \begin{tikzpicture}
            \begin{axis}[
                width=4.8cm,
                height=4.8cm,
                xlabel={Disruption cost $\alpha$},
                ylabel={Cost reduction ($\%$)},
                xtick={0, 0.1, 0.2, 0.3, 0.4},
                ymin=0,
                ymax=100,
                grid=both
            ]
    
            \pgfplotstableread{
                x       y
                0       67
                0.1    66.67
                0.2     65
                0.3    54
                0.4       46.7
            } \regionTwo;

            \pgfplotstableread{
                x       y
                0       66.33
                0.1    63.17
                0.2     51.33
                0.3    37.17
                0.4       33.17
            } \regionFive;
    
            \addplot table[x=x, y=y] {\regionTwo};
            \addlegendentry{R=2}
    
            \addplot[style=densely dashed,color=blue,mark=square*,mark options=solid] table[x=x, y=y] {\regionFive};
            \addlegendentry{R=5}

            \end{axis}
        \end{tikzpicture}
        \caption{Deployment cost}
        \label{subfig:opt_sol_cost}
    \end{subfigure}%
    \begin{subfigure}[b]{0.25\textwidth}
        \begin{tikzpicture}
            \begin{axis}[
                width=4.8cm,
                height=4.8cm,
                xlabel={Disruption cost $\alpha$},
                ylabel={No. of microservices},
                xtick={0, 0.1, 0.2, 0.3, 0.4},
                ymin=0,
                ymax=16,
                xmin=-0.05,
                xmax=0.45,
                grid=both
            ]

            \pgfplotstableread{
                x       y
                0       29.8
                0.1    10.6
                0.2     7.4
                0.3    6.6
                0.4       5.6
            } \regionTwo;
    
            \pgfplotstableread{
                x       y
                0       28
                0.1    10
                0.2     6.7
                0.3    3.8
                0.4      3.0
            } \regionFive;
    
            \pgfplotstableread{
                x       y
                0       0.12
                0.1    0.14
                0.2    0.18
                0.3    0.22
                0.4       0.28
            } \heu;
    
            \addplot table[x=x, y=y] {\regionTwo};
            \addlegendentry{R=2}
    
            \addplot[style=densely dashed,color=blue,mark=square*,mark options=solid] table[x=x, y=y] {\regionFive};
            \addlegendentry{R=5}       
            \end{axis}
        \end{tikzpicture}
         \caption{Rescheduled microservices}
        \label{subfig:opt_sol_pods}
    \end{subfigure}%
    \begin{subfigure}[b]{0.25\textwidth}
        \begin{tikzpicture}
            \begin{axis}[
                width=4.8cm,
                height=4.8cm,
                xlabel={Disruption cost $\alpha$},
                ylabel={Time (s)},
                xtick={0, 0.1, 0.2, 0.3, 0.4},
                ymin=0,
                ymax=65,
                grid=both
            ]

            \pgfplotstableread{
                x       y
                0      55
                0.1    15.5
                0.2     6.8
                0.3    6.1
                0.4       5
            } \regionTwo;

            \pgfplotstableread{
                x       y
                0       51 
                0.1     7.5
                0.2     5.5
                0.3     5
                0.4     5
            } \regionFive;
    
            \addplot table[x=x, y=y] {\regionTwo};
            \addlegendentry{R=2}
    
            \addplot[style=densely dashed,color=blue,mark=square*,mark options=solid]  table[x=x, y=y] {\regionFive};
            \addlegendentry{R=5}        
            \end{axis}
        \end{tikzpicture}
        \caption{Rescheduling duration}
        \label{subfig:opt_sol_slots}
    \end{subfigure}%
    \begin{subfigure}[b]{0.25\textwidth}
        \begin{tikzpicture}
            \begin{axis}[
                width=4.8cm,
                height=4.8cm,
                xlabel={Cost objective ($\%$)},
                ylabel={Disruption objective ($\%$)},
                xtick={0, 25, 40, 55, 70},
                xmin=20,
                xmax=75,
                ymin=0,
                ymax=65,
                grid=both
            ]
    
            \pgfplotstableread{
                y       x
                63.4      33
                21.4   33.33
                13.8        35
                11.3        46
                10.3     53.3
                0        100
            } \regionTwo;

            \pgfplotstableread{
                y       x
                59.7   33.67
                18.6    36.8
                12.2     48.7
                7.5     62.8
                6.25     66.8
                0        100
            } \regionFive;
    
            \addplot[const plot,style=solid,color=blue] table[x=x, y=y] {\regionTwo};
            \addlegendentry{R=2}
    
            \addplot[const plot, style=densely dashed,color=blue] table[x=x, y=y] {\regionFive};
            \addlegendentry{R=5}

            \end{axis}
        \end{tikzpicture}
        \caption{Pareto frontier}
        \label{subfig:opt_pareto}
    \end{subfigure}
    \caption{Optimal microservice rescheduling computed with different values of disruption cost $\alpha$ when the number of cloud regions is $R=2$ and $R=5$. The number of microservices is $M=32$. The number of nodes is $N=10$.}
    \label{fig:opt_sol}
\end{figure*}
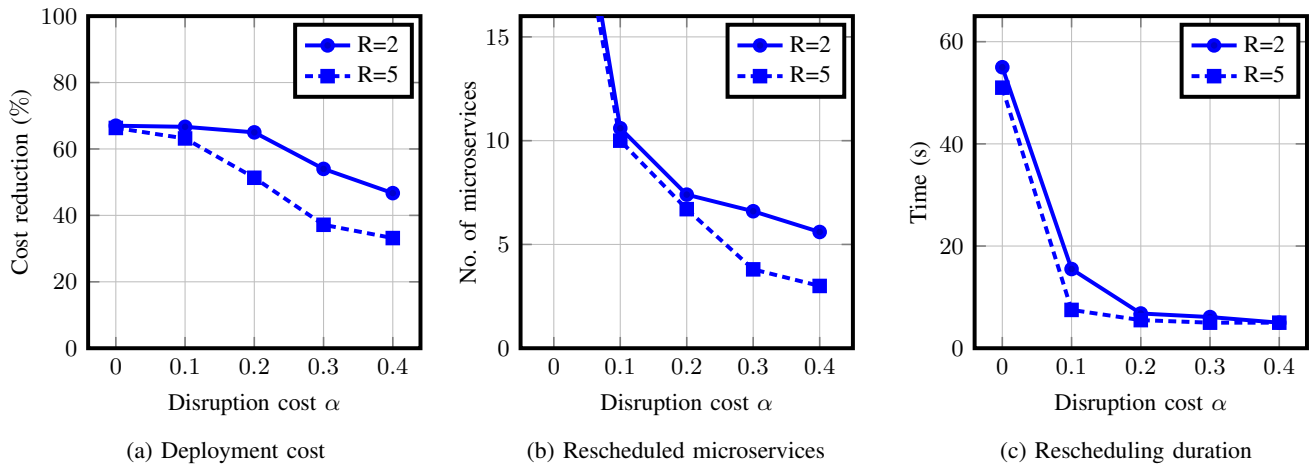

In Fig.~\ref{fig:architecture}, we provide an overview of the system architecture used to execute the experiments along with the typical workflow, that consists of the following steps: 
\begin{enumerate}
    \item The \texttt{Workload generator} provides a feasible deployment configuration $S_A$ consisting of a set of microservice-based applications with predefined resource requirements. This configuration is deployed using the default Kubernetes orchestrator (0.)
    \item The \texttt{Service provider} triggers the rescheduling of microservices due to a low resource utilization efficiency of the current microservice configuration  (1.)
    \item A custom \texttt{Kubectl plugin}, which is the component allowing to communicate with Kubernetes control plane, collects the status of the virtual infrastructure and the workload deployed and invokes the \texttt{Algorithm} module (2.)
    \item The \texttt{Algorithm} module implements the Optimal and Heuristic schemes and outputs the new placement for each microservice as well as the related rescheduling time slot in which to execute the node reassignment.  This information is sent back to the \texttt{Kubectl plugin}  in the form of placements and ranking scores for each microservice (3.)
    \item The \texttt{K8s Scheduler} receives the ordered list of microservices to be redeployed. Two custom plugins have been added to the \texttt{K8s Scheduler} at the extension points \texttt{Pre-Enqueue} and \texttt{Score}: the first one allows to order the microservices according to what requested by the \texttt{Algorithm}, while the second one applies the scores (4.). 
    \item Finally, the planned rescheduling of microservices is executed on the selected nodes by the \texttt{K8s Scheduler} (5.).
\end{enumerate}

\begin{table}[t] 
\centering
\caption{Simulation parameters}\label{tab1}
\begin{tabular}{lc}
\toprule
Number of applications & \{8, 100\} \\
Microservices per application & 4 \\
Number of cloud regions & \{2, 5\} \\
Inter-region network latency & 50ms \\
Number of virtual nodes & \{10, 100\} \\
Normalized cost per multi-cloud node  & $\{0,1,...,4\}$ \\
Number of vCPU & 8 cores \\
Available RAM & 32 GB \\
Microservice CPU request & $[0.5,2.5]$ vCPU \\
Microservice RAM request & $[0.2,0.3]$ GB\\
\bottomrule
\end{tabular}
\end{table}

We generate the virtual infrastructure as virtual nodes of equal capacity $C_n$ grouped in $R$ differently-priced cloud regions having a linearly increasing cost per unit of resources (for example, assuming R=3, each region would have $p_n =0$ for region 1, $p_n =1$ for region 2 and $p_n=2$ for region 3). 
The feasible deployment of microservice $S_A$ is computed following three main steps. First,  we randomly generated the resource requirements for each microservice according to a uniform distribution of $[0.5,2.5]$ for CPU resource and $[0.2,0.3]$ for RAM resources. We employed these values by analyzing  Alibaba cluster traces, which are a publicly available collection of statistics of real microservices deployed on Alibaba production infrastructure \cite{everman2021improving}. Then, to artificially produce a fragmented resource utilization, we randomly allocated the microservices among the various virtual nodes enforcing an average resource occupation of $70\%$ with respect to the maximum node capacity. Finally, to ensure that all QoS requirements were met prior to the re-orchestration process, we retrospectively defined the co-location constraints $\ell^{(n,n')}_{a,m,m'}$ for the deployed microservices. These requirements were computed by randomly sampling a maximum latency within the range of $[20 \text{ms},100\text{ms}]$. The lower bound represents latency-critical communication flows, while the upper bound corresponds to latency-agnostic flows. We resume the main configuration parameters in Table \ref{tab1}.
We generated 50 feasible deployment configuration $S_A$ for each combination of number of microservices and virtual nodes grouped in different cloud regions and we averaged the results obtained with the various schemes. The statistical reliability of the results has been assessed by computing the 95\% confidence intervals. However, they are not plotted, as their narrow range made them visually negligible. In the worst case, the interval spanned only 3\% of the corresponding average value.

\subsection{Optimal solution analysis}

In Fig.  \ref{fig:opt_sol}, we provide a sensitivity analysis of the performance of the optimal solution for different values of the disruption cost $\alpha$. Moreover, we also show the Pareto frontier in order to better highlight the trade-off between deployment cost and service disruption objectives. We consider a simple scenario with $A=8$ applications of 4 microservices each, that are allocated on $N=10$ virtual nodes equally distributed in $R=2$ region and $R=5$ regions, respectively. In detail, in Fig. \ref{subfig:opt_sol_cost}, the increase of the service disruption cost decreases the cost minimization gain. This trend stems from an increase in the penalty magnitude applied to the objective function \eqref{eq_obj_func} when rescheduling microservices across multiple time slots.  This dynamic incentives a rescheduling solution that greatly limits the number of reallocated microservices and that employs few rescheduling time slots, thus decreasing the rescheduling duration, as shown in Fig. \ref{subfig:opt_sol_pods} and in Fig. \ref{subfig:opt_sol_slots}, respectively. This effect is more exacerbated when the number of cloud regions is $R=5$ compared to $R=2$ since the more challenging QoS requirements restricts the possible rescheduling options. As a matter of fact, the repacking of microservices in cheaper cloud regions is exclusively performed only if the related nodes have enough resources to fully accommodate microservices with co-location requirements. In Fig. \ref{subfig:opt_pareto}, we provide a holistic view of this behavior by computing the minimum level of service disruption (expressed by the second term of the objective function \eqref{eq_obj_func})  generated to obtain a target deployment cost. Both objectives have been normalized with respect to their worst-case performance: the highest service disruption corresponds to the sequential rescheduling of all microservices, whereas the highest deployment cost corresponds to the original deployment cost prior to the re-orchestration. Although the service disruption becomes rapidly more severe as the deployment cost decreases,  this trend is moderate for less stringent cost targets and can  be approximated to the scenario obtained with $\alpha = 0.1$. For this reason, we use this configuration to obtain a balanced optimal rescheduling solution that  minimizes the service disruption  without significantly deteriorating the cost minimization performance. 
\subsection{Heuristic solution analysis}

\pgfplotsset{every axis/.append style={font=\small}}
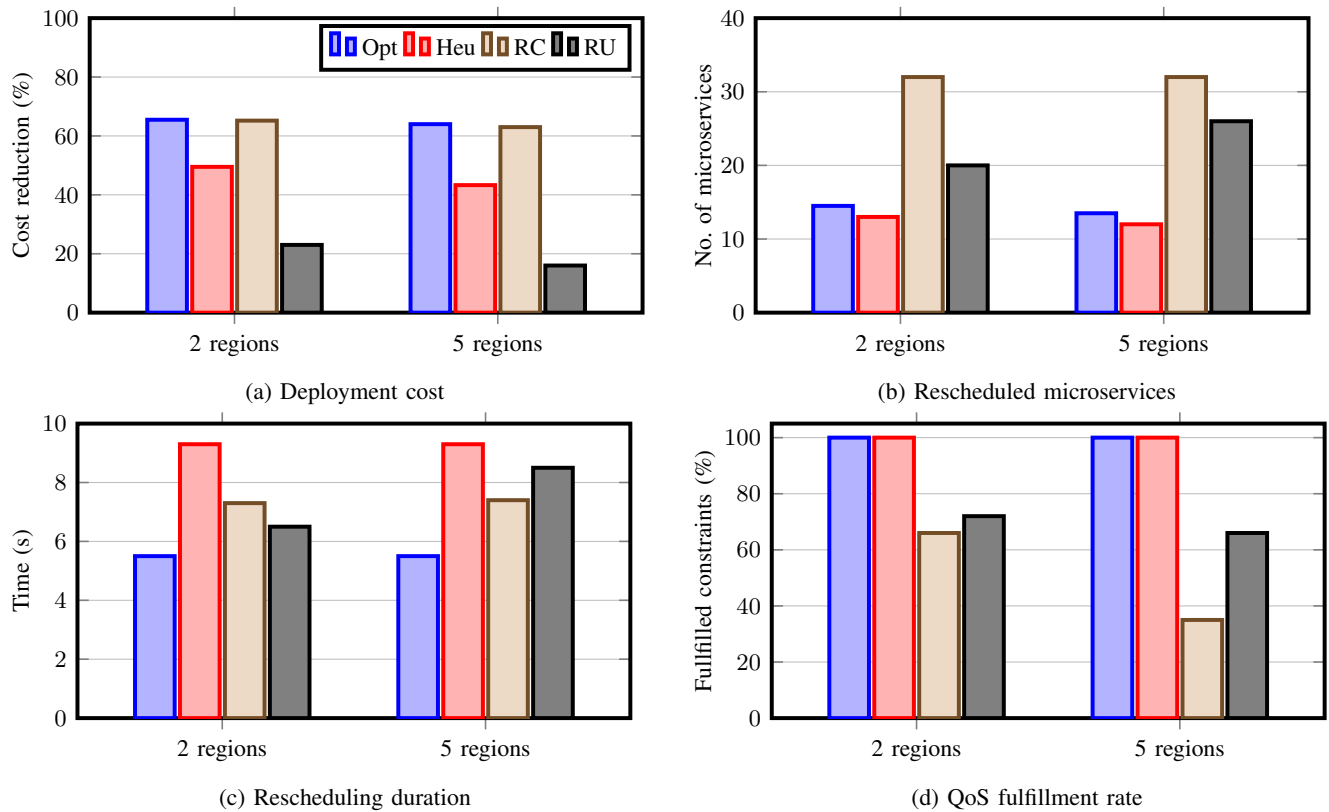
\begin{figure*}
    \centering
    \begin{subfigure}[b]{0.25\textwidth}
        \begin{tikzpicture}
        \begin{axis}[
                    width=4.8cm,
                    height=4.8cm,
                    ybar,
                    legend style={legend columns=1},
                        legend image code/.code={
        \draw[mark=*, mark options={scale=1.2}] (0cm,-0.1cm) -- (0cm,0.1cm);
    },
                    ylabel={Cost reduction (\%)},
                    symbolic x coords={dummy},
                    xtick=\empty,       
                    enlarge x limits=0.4,       
                    bar width=10pt,
                    x=1.5cm,
                    ymin=0,
                    ymax=100,
                    ymajorgrids=true
                ]
        \addplot coordinates {(dummy,65.5)};
        \addplot coordinates {(dummy,63.6)};
        \addplot coordinates {(dummy,65.2)};
        \addplot coordinates {(dummy,17.6)};
        \addplot coordinates {(dummy,23)};
          \legend{Opt,Heu,B-RC,B-RU,K8S}
        \end{axis}
        \end{tikzpicture}
        \caption{Deployment cost}
         \label{subfig:heu_sol_cost}
    \end{subfigure}%
    \begin{subfigure}[b]{0.25\textwidth}
                \begin{tikzpicture}
        \begin{axis}[
            width=4.8cm,
            height=4.8cm,
            ybar,
            enlarge x limits=0.2,
            legend style={legend columns=1},
            ylabel={No. of microservices},
           symbolic x coords={2 regions},
            xtick=\empty,
            bar width=10pt,
            x=1.5cm,
            ymin=0,
            ymax=40,
            ymajorgrids=true
        ]
        \addplot coordinates {(2 regions,14.5) };
        \addplot coordinates {(2 regions,15.6) };
        \addplot coordinates {(2 regions, 32) };
        \addplot coordinates {(2 regions, 32) };
        \addplot coordinates {(2 regions, 20)  };
        \end{axis}
        \end{tikzpicture}
        \caption{Rescheduled microservices}
        \label{subfig:heu_sol_pods}
    \end{subfigure}%
    \begin{subfigure}[b]{0.25\textwidth}
                \begin{tikzpicture}
        \begin{axis}[
            width=4.8cm,
            height=4.8cm,
            ybar,
            enlarge x limits=0.2,
            legend style={legend columns=-1},
            ylabel={Time (s)},
            symbolic x coords={2 regions},
            xtick=\empty,
            bar width=10pt,
            x=1.5cm,
            ymin=0,
            ymax=10,
            ymajorgrids=true
        ]
        \addplot coordinates {(2 regions,5.5)  };
        \addplot coordinates {(2 regions,9.3) };
        \addplot coordinates {(2 regions,7.3) };
        \addplot coordinates {(2 regions,9.4) };
        \addplot coordinates {(2 regions,6.5)  };
        \end{axis}
        \end{tikzpicture}
        \caption{Rescheduling duration}
        \label{subfig:heu_sol_slots}
    \end{subfigure}%
    \begin{subfigure}[b]{0.25\textwidth}
                \begin{tikzpicture}
        \begin{axis}[
            width=4.8cm,
            height=4.8cm,
            ybar,
            enlarge x limits=0.2,
            legend style={legend columns=-1},
            ylabel={Fullfilled constraints (\%)},
           symbolic x coords={2 regions},
            xtick=\empty,
            ytick= {0,20,40,60,80,100},
            bar width=10pt,
            x=1.5cm,
            ymin=0,
            ymax=105,
            ymajorgrids=true
        ]
        \addplot coordinates {(2 regions,100) };
        \addplot coordinates {(2 regions,100) };
        \addplot coordinates {(2 regions,100) };
        \addplot coordinates {(2 regions,72) };
        \addplot coordinates {(2 regions,64) };
        \end{axis}
        \end{tikzpicture}
        \caption{QoS fulfillment rate}
        \label{subfig:heu_sol_qos}
    \end{subfigure}
    \caption{ Microservice rescheduling performance obtained when the number of cloud regions is $R \!= 2$, the number of microservices is $M\!=\!32$ and the number of nodes is $N \!=\! 10$. The Optimal scheme is computed with $\alpha = 0.1$.}
    \label{fig:heu_sol}
\end{figure*}

In Fig. \ref{fig:heu_sol}, we extend the performance analysis by including the Heuristic scheme and we compare the results obtained with respect to the benchmark schemes B-RC, B-RU and K8S. In Fig. \ref{subfig:heu_sol_cost}, we observe that the Heuristic scheme achieve a cost reduction close to the Optimal scheme while outperforming the default K8S scheduler implementation. Similarly,  the B-RC scheme provides comparable performance which is however achieved by ignoring the service continuity degradation caused by the \textit{Recreate} deployment strategy. The impact of this procedure on costs is visible when observing the performance of B-RU which instead employs a \textit{Rolling Update} deployment strategy. The considerable performance degradation is explained by the fact that B-RU ignores the coexistence of old and new microservice replicas when  computing the rescheduling policy. This limitation causes a mismatch between the planned microservice allocation and the actual executed one (see the example in Fig. \ref{fig:resch_example}) which results in some microservice not being rescheduled on the predetermined destination node. In contrast, since B-RC redeploys all applications from scratch, it can achieve better repacking performance because nodes are initially empty and resources can be used more efficiently. The downside of this approach, as shown in Fig. \ref{subfig:heu_sol_pods}, produces the highest number of rescheduled microservices that are in fact first terminated before being reallocated to the new destination nodes. Differently, the heuristic scheme reschedules a significantly lower number of microservices, which is comparable to the optimal solution, by selectively repacking only microservices that offer the largest cost reduction based on the available resources in the current rescheduling time slot. This strategy also reduces the number of rescheduled microservices compared to K8S, which ignores the increase in resource contention caused by the coexistence of old and new microservice instances within the same rescheduling slot. Overlooking this effect produces an uncoordinated rescheduling of microservices that does not only degrade cost performance but also extends the rescheduling duration as depicted in Fig. \ref{subfig:heu_sol_slots}.
Furthermore, as shown in Fig. \ref{subfig:heu_sol_qos}, the heuristic scheme fully preserves the QoS performance as it achieves the maximum QoS fulfillment rate like the optimal solution. Likewise, B-RC achieve optimal QoS performance due to its rescheduling strategy that prioritizes the allocation of microservices with co-location requirements in order to maximize the chance to successfully satisfy their constraints. In contrast, the configuration using the rolling update strategy, i.e. B-RU, produces worse results which is again explained by the mismatch between the planned rescheduling policy and executed allocation splitting microservices in different cloud regions. Similarly, K8S produces modest QoS fulfillment performance as it tries to simultaneously achieve cost efficiency and QoS preservation at the same time by poorly combining the contribution of $NodeAffinity$ and $PodAffinity$ values.

\pgfplotsset{every axis/.append style={line width=1.5pt,tick style={line width=0.7pt}}}
\pgfplotsset{every axis/.append style={font=\small}}
\pgfplotsset{every axis y label/.append style={yshift=-2pt}}
\pgfplotmarksize=2.5pt
\begin{figure*}[t]
    \centering
    \begin{subfigure}[b]{0.25\textwidth}
        \begin{tikzpicture}
               \begin{axis}[
                width=4.8cm,
                height=4.8cm,
                xlabel={Cloud regions},
                ylabel={Cost reduction ($\%$)},
                xtick={2, 3, 4, 5},
                ytick={0,20,40,60,80,100},
                ymin=0,
                ymax=90,
                grid=both,
                cycle multiindex* list={{red,brown,black,violet}\nextlist mark list*}
            ]
            
            \pgfplotstableread{
                x       y       yerr
                2       55.2      0.5
                3       53.1      2
                4       51.4      4
                5       47.5      3
            } \Heu;

            \pgfplotstableread{
                x       y
                2       56.6
                3       56
                4       54
                5       52
            } \Recreate;
            
            \pgfplotstableread{
                x       y
                2       17
                3       15
                4       14.2
                5       14.1
            } \Rolling;
            
            \pgfplotstableread{
                x       y
                2       56.5
                3       55.9
                4       54.4
                5       52.4
            } \BRC;
            
            \pgfplotstableread{
                x       y
                2       1.2
                3       1.4
                4       1.4
                5       2
            } \BRU;

            \addplot table[x=x, y=y] {\Heu};
            \addplot table[x=x, y=y] {\BRC};
            \addplot table[x=x, y=y] {\BRU};
            \addplot table[x=x, y=y] {\Rolling};
            
            \end{axis}
        \end{tikzpicture}
        \caption{Deployment cost}
        \label{subfig:heu_sol_large_cost}
    \end{subfigure}%
    \begin{subfigure}[b]{0.25\textwidth}
        \begin{tikzpicture}
            \begin{axis}[
                width=4.8cm,
                height=4.8cm,
                xlabel={Cloud regions},
                ylabel={No. of microservices},
                xtick={2, 3, 4, 5},
                ytick={0,100, 200, 300, 400},
                ymin=0,
                ymax=420,
                grid=both,
                cycle multiindex* list={{red,brown,black,violet}\nextlist mark list*}
            ]

            \pgfplotstableread{
                x       y
                2       125
                3       129.5
                4       142.7
                5       141.2
            } \Heu;

            \pgfplotstableread{
                x       y
                2       400
                3       400
                4       400
                5       400
            } \Recreate;

            \pgfplotstableread{
                x       y
                2       200
                3       268
                4       300
                5       320
            } \Rolling;

            \pgfplotstableread{
                x       y
                2       400
                3       400
                4       400
                5       400
            } \BRC;

            \pgfplotstableread{
                x       y
                2       400
                3       400
                4       400
                5       400
            } \BRU;

            \addplot table[x=x, y=y] {\Heu};
    

            \addplot  table[x=x, y=y] {\BRC};
            \addplot  table[x=x, y=y] {\BRU};
            \addplot  table[x=x, y=y] {\Rolling};
            
            \end{axis}
        \end{tikzpicture}
         \caption{Rescheduled microservices}
        \label{subfig:heu_sol_large_pods}
    \end{subfigure}%
    \begin{subfigure}[b]{0.25\textwidth}
        \begin{tikzpicture}
            \begin{axis}[
                width=4.8cm,
                height=4.8cm,
                xlabel={Cloud regions},
                ylabel={Time (s)},
                legend style={at={(0.05,0.97)}, anchor=north west},
                xtick={2, 3, 4, 5},
                ymin=0,
                ymax=300,
                grid=both,
                cycle multiindex* list={{red,brown,black,violet}\nextlist mark list*}
            ]

            \pgfplotstableread{
                x       y
                2       65
                3       63
                4       67.9
                5       68.1
            } \Heu;

            \pgfplotstableread{
                x       y
                2       77
                3       104
                4       116
                5       126
            } \Rolling;

            \pgfplotstableread{
                x       y
                2       215
                3       215.5
                4       217
                5       219
            } \BRC;

            \pgfplotstableread{
                x       y
                2       193.4
                3       193.8
                4       192.9
                5       193.2
            } \BRU;

            \addplot table[x=x, y=y] {\Heu};
    
            \addplot  table[x=x, y=y] {\BRC};

            \addplot  table[x=x, y=y] {\BRU};

            \addplot  table[x=x, y=y] {\Rolling};

            \end{axis}
        \end{tikzpicture}
        \caption{Rescheduling duration}
        \label{subfig:heu_sol_large_slots}
    \end{subfigure}%
    \begin{subfigure}[b]{0.25\textwidth}
        \begin{tikzpicture}
            \begin{axis}[
                width=4.8cm,
                height=4.8cm,
                xlabel={Cloud regions},
                ylabel={Fulfilled constraints (\%)},
                legend style={at={(0.02,0.03)}, anchor=south west},
                xtick={2, 3, 4, 5},
                ytick={0,20, 40, 60, 80, 100},
                ymin=0,
                ymax=110,
                grid=both,
                cycle multiindex* list={{red,brown,black,violet}\nextlist mark list*}
            ]

            \pgfplotstableread{
                x       y
                2       100
                3       100
                4       100
                5       100
            } \Heu;

            \pgfplotstableread{
                x       y
                2       26.2
                3       9.7
                4       4
                5       2
            } \Recreate;

            \pgfplotstableread{
                x       y
                2       79.3
                3       68.9
                4       56.6
                5       45.4
            } \Rolling;

            \pgfplotstableread{
                x       y
                2       100
                3       100
                4       94.7
                5       93.5
            } \BRC;

            \pgfplotstableread{
                x       y
                2       86
                3       71.7
                4       69.3
                5       68.3
            } \BRU;

            \addplot table[x=x, y=y] {\Heu};
            \addlegendentry{Heu}
    

            \addplot  table[x=x, y=y] {\BRC};
            \addlegendentry{B-RC}
            \addplot  table[x=x, y=y] {\BRU};
            \addlegendentry{B-RU}
            \addplot  table[x=x, y=y] {\Rolling};
            \addlegendentry{K8S}

            \end{axis}
        \end{tikzpicture}
        \caption{QoS fulfillment rate}
        \label{subfig:heu_sol_large_qos}
    \end{subfigure}
    \caption{Microservice rescheduling performance as the number of cloud region increases from $R\!=2$ to $R\!=5$. The number of microservices is $M\!=400$. The total number of nodes is $N\! = 100$.}
    \label{fig:heu_sol_large}
\end{figure*}
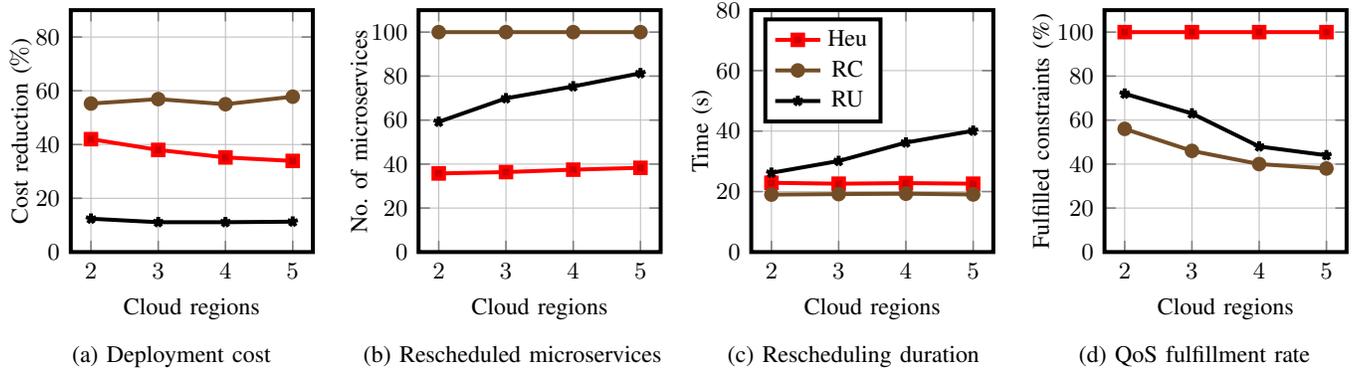

\pgfplotsset{every axis/.append style={line width=1.5pt,tick style={line width=0.7pt}}}
\pgfplotsset{every axis/.append style={font=\small}}
\pgfplotsset{every axis y label/.append style={yshift=-2pt}}
\pgfplotmarksize=2.5pt
\begin{figure*}[t]
    \centering
    \begin{subfigure}[b]{0.25\textwidth}
        \begin{tikzpicture}
            \begin{axis}[
                width=4.8cm,
                height=4.8cm,
                xlabel={Average node load (\%)},
                ylabel={Cost reduction ($\%$)},
                xtick={30, 50, 70, 90},
                xmin= 25,
                ytick={0,20, 40, 60, 80, 100},
                ymin=0,
                ymax=110,
                grid=both,         
                cycle multiindex* list={{red,brown,black,violet}\nextlist mark list*}
            ]
    
            \pgfplotstableread{
                x       y
                30       93.1
                50       75.1
                70       47.5
                90       16.6
            } \Heu;

            \pgfplotstableread{
                x       y
                30      93.5
                50    41.5
                70    14.12
                90    0.29
            } \Rolling;

            \pgfplotstableread{
                x       y
                30     94.5
                50    78.6
                70    52.4
                90    21.1
            } \BRC;

            \pgfplotstableread{
                x       y
                30      93.1
                50    55
                70    2.1
                90    0.39
            } \BRU;

            \addplot table[x=x, y=y] {\Heu};
    
            \addplot  table[x=x, y=y] {\BRC};
            \addplot  table[x=x, y=y] {\BRU};
            \addplot  table[x=x, y=y] {\Rolling};
             
            \end{axis}
        \end{tikzpicture}
        \caption{Deployment cost}
        \label{subfig:heu_sol_variable_load_cost}
    \end{subfigure}%
    \begin{subfigure}[b]{0.25\textwidth}
        \begin{tikzpicture}
            \begin{axis}[
                width=4.8cm,
                height=4.8cm,
                xlabel={Average node load (\%)},
                ylabel={No. of microservices},
                ylabel style={yshift=1mm},
                xtick={30, 50, 70, 90},
                ytick={0,125, 250, 375, 500},
                ymin=0,
                ymax=520,
                grid=both,
                cycle multiindex* list={{red,brown,black,violet}\nextlist mark list*}
            ]

            \pgfplotstableread{
                x       y
                30       119.8
                50       150.8
                70       141.18
                90       91.7
            } \Heu;

            \pgfplotstableread{
                x       y
                30       139
                50       228
                70       320
                90       401
            } \Rolling;

            \pgfplotstableread{
                x       y
                30       170
                50       285
                70       400
                90       500
            } \BRC;
            
            \pgfplotstableread{
                x       y
                30       170
                50       285
                70       400
                90       500
            } \BRU;

            \addplot table[x=x, y=y] {\Heu};   
            \addplot  table[x=x, y=y] {\BRU};
            \addplot  table[x=x, y=y] {\BRU};
            \addplot  table[x=x, y=y] {\Rolling};
            
            \end{axis}
        \end{tikzpicture}
         \caption{Rescheduled microservices}
        \label{subfig:heu_sol_variable_load_pods}
    \end{subfigure}%
    \begin{subfigure}[b]{0.25\textwidth}
        \begin{tikzpicture}
            \begin{axis}[
                width=4.8cm,
                height=4.8cm,
                xlabel={Average node load (\%)},
                ylabel={Time (s)},
                ylabel style={yshift=1mm},
                legend style={at={(0.05,0.97)}, anchor=north west},
                xtick={30, 50, 70, 90},
                ymin=0,
                ymax=320,
                grid=both,
                cycle multiindex* list={{red,brown,black,violet}\nextlist mark list*}
            ]

            \pgfplotstableread{
                x       y
                30       53
                50       71.3
                70       67.4
                90       60.9
            } \Heu;

            \pgfplotstableread{
                x       y
                30       52.5
                50       83
                70       125.5
                90       165.6
            } \Rolling;

            \pgfplotstableread{
                x       y
                30       70.8
                50       130.4
                70       221
                90       308
            } \BRC;
            
            \pgfplotstableread{
                x       y
                30       71
                50       121
                70       195
                90       280
            } \BRU;

            \addplot table[x=x, y=y] {\Heu};
            \addlegendentry{Heu}
    

            \addplot  table[x=x, y=y] {\BRC};
            \addlegendentry{B-RC}

            \addplot  table[x=x, y=y] {\BRU};
            \addlegendentry{B-RU}

            \addplot  table[x=x, y=y] {\Rolling};
            \addlegendentry{K8S}

            \end{axis}
        \end{tikzpicture}
        \caption{Rescheduling duration}
        \label{subfig:heu_sol_variable_load_slots}
    \end{subfigure}%
    \begin{subfigure}[b]{0.25\textwidth}
        \begin{tikzpicture}
            \begin{axis}[
                width=4.8cm,
                height=4.8cm,
                xlabel={Average node load (\%)},
                ylabel={Fulfilled constraints (\%)},
                xtick={30, 50, 70, 90},
                ytick={0,20, 40, 60, 80, 100},
                ymin=0,
                ymax=110,
                grid=both,
                cycle multiindex* list={{red,brown,black,violet}\nextlist mark list*}
            ]

            \pgfplotstableread{
                x       y
                30       100
                50       100
                70       100
                90       100
            } \Heu;

            \pgfplotstableread{
                x       y
                30       89.8
                50       69.9
                70       45.4
                90       24.9
            } \Rolling;

            \pgfplotstableread{
                x       y
                30       100
                50       97
                70       93.6
                90       94
            } \BRC;

            \pgfplotstableread{
                x       y
                30       88
                50       80
                70       68.3
                90       29.3
            } \BRU;

            \addplot table[x=x, y=y] {\Heu};
    
            \addplot  table[x=x, y=y] {\BRC};

            \addplot  table[x=x, y=y] {\BRU};
            \addplot  table[x=x, y=y] {\Rolling};

            \end{axis}
        \end{tikzpicture}
        \caption{QoS fulfillment rate}
        \label{subfig:heu_sol_variable_load_qos}
    \end{subfigure}
    \caption{Microservice rescheduling performance as the average resource occupation per node is increased from 30\% to 90\%. The total number of microservices depends on the target resource load. The total number of nodes is $N\! = 100$.}
    \label{fig:heu_sol_variable_load}
\end{figure*}

We extend the evaluation on a larger virtual infrastructure and microservice deployment configuration in order to showcase performance on a more realistic multi-cloud scenario. Due to the high computational complexity of the Optimal solution, we only show the results of the Heuristic scheme. In Fig. \ref{fig:heu_sol_large}, we show the overall performance achieved by the heuristic scheme for an increasing number of multi-cloud regions starting from $R=2$ up to $R=5$ when total number of applications is $A=100$ (i.e., $M=400$ microservices) and the total number of nodes $N = 100$. In general, the same observations made for the scenario with $M=32$ hold also for this case. The heuristic scheme ensures the most efficient trade-off in terms of cost reduction, service disruption, and QoS preservation. In particular, as shown in Fig. \ref{subfig:heu_sol_large_cost}, the cost minimization performance achieved by the heuristic scheme decreases with the number of regions. This is caused by a more restrictive rescheduling of microservices, which is needed to avoid the separation of co-located microservices in different cloud regions. As shown in Fig. \ref{subfig:heu_sol_large_qos}, this conservative approach allows the heuristic scheme to maximize the QoS fulfillment rate regardless of the number of cloud regions. Moreover, although B-RC achieves slightly better cost performance, it introduces a very high service disruption since all microservices are terminated before being rescheduled. On the other hand, the rolling update strategy employed by B-RU is ineffective at reducing costs, primarily due to a significant mismatch between its rescheduling policy and the actual allocation, which results in a largely random distribution of microservices across nodes. A similar trend is observed using K8s that suffers from the limitation of combining multiple objectives using different affinity-based rules. Furthermore, the larger configuration of microservices and nodes also amplifies the resource contention issue that impacts more severely the rescheduling duration of the various benchmark schemes, as shown in Fig. \ref{subfig:heu_sol_large_slots}. Conversely, it moderately affects the heuristic scheme, confirming its capability to effectively parallelize the microservice rescheduling process. In addition, as shown in Fig. \ref{subfig:heu_sol_large_pods}, the heuristic scheme reschedules a similar number of microservices for each multi-cloud configuration, as it repacks them based on a combination of resource availability and potential cost reduction. This approach inherently limits the number of rescheduling opportunities. In contrast, K8s repacks  microservices allocated on the most expensive cloud regions. As a result, from a service disruption standpoint, K8s produces progressively poorer performance compared to the heuristic scheme as the number of multi-cloud region increases. 

Beside a variable number of cloud regions, in Fig. \ref{fig:heu_sol_variable_load} we also evaluate the performance for a variable number of microservices. This analysis provides insights on the performance of the heuristic algorithm in heterogeneous load conditions and allows to showcase its behavior under different workload regimes. In detail, in low load scenarios, the various schemes have similar performance due to the low resource contention that trivializes the microservice rescheduling computation. As a matter of fact, any rescheduling strategy that repacks microservices on the cheapest nodes can provide considerable cost savings.   However, as the load increases, the possible rescheduling options become more restrictive and heavily affect the cost performance of B-RU and K8s which fails in reducing costs when the average resource load is at 90\%,  as shown in \ref{subfig:heu_sol_variable_load_cost}. Conversely, the heuristic scheme achieves notable cost savings even under heavy load conditions by preventing resource contention. This is made possible by its rescheduling strategy, which dynamically determines the number of microservices to be rescheduled based on the available resources, as illustrated \ref{subfig:heu_sol_variable_load_pods}. This behavior reduces the rescheduling duration, which is significantly lower than the other schemes that reschedule most microservices regardless of the load condition as depicted in Fig. \ref{subfig:heu_sol_variable_load_slots}. Moreover, in Fig. \ref{subfig:heu_sol_variable_load_qos}, we observe that the mandatory enforcement of the co-location requirements performed by the heuristic scheme ensures maximum fulfillment rate. In contrast, although the B-RC employs a prioritization strategy to promote QoS enforcement, its performance deteriorates in heavy load scenarios due to restrictive rescheduling configuration that makes the computation of a QoS-compliant rescheduling solution more challenging.

\subsection{Computational complexity analysis}

\pgfplotsset{every axis/.append style={line width=1.5pt,tick style={line width=0.6pt}}}
\pgfplotsset{every axis/.append style={font=\small}}
\pgfplotsset{every axis y label/.append style={yshift=-2pt}}
\pgfplotmarksize=2pt
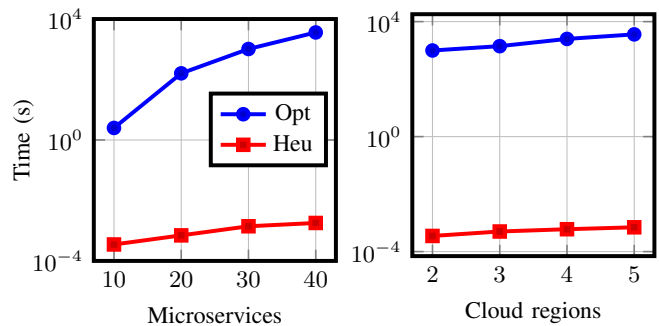
\begin{figure}[t]
    \centering
    \begin{subfigure}[b]{0.24\textwidth}
        \begin{tikzpicture}
            \begin{axis}[
                width=4.8cm,
                height=4.8cm,
                xlabel={Microservices},
                ylabel={Time (s)},
                ymode=log,
                legend style={at={(0.475,0.7)}, anchor=north west},
                ymin=0.0001,
                ymax=10000,
                grid=both,
                ]
    
            \pgfplotstableread{
                x       y
                10       2.5
                20    160
                30     1020
                40    3600
            } \Opt;

            \pgfplotstableread{
                x       y
                10       0.0004
                20    0.0007
                30     0.0014
                40    0.0018
            } \Heu;

            \addplot table[x=x, y=y] {\Opt};
            \addlegendentry{Opt}
    
            \addplot  table[x=x, y=y] {\Heu};
            \addlegendentry{Heu}    

            \end{axis}
        \end{tikzpicture}
        \caption{Resolution time for $R\!=\!5$.}
        \label{subfig:time_opt_heu}
    \end{subfigure}\hspace{0.01cm}
    \begin{subfigure}[b]{0.24\textwidth}
        \begin{tikzpicture}
            \begin{axis}[
                width=4.8cm,
                height=4.8cm,
                xlabel={Cloud regions},
                legend style={at={(0.02,0.98)}, anchor=north west},
                xtick={2, 3, 4, 5},
                ymode=log,
                grid=both
            ]

            \pgfplotstableread{
                x       y
                2       990
                3    1400
                4     2500
                5    3600
            } \Opt;

            \pgfplotstableread{
                x       y
                2       0.00035
                3    0.0005
                4     0.0006
                5    0.0007
            } \Heu;

            \addplot table[x=x, y=y] {\Opt};
    
            \addplot  table[x=x, y=y] {\Heu};

            \end{axis}
        \end{tikzpicture}
          \caption{Resolution time for $M\!=\!40$.}
        \label{subfig:time_heu_k8s}
    \end{subfigure}%
    \caption{Computational performance of the Optimal and Heuristic schemes obtained with different configurations of microservices and cloud regions. The number of nodes is $N=10$.}
    \label{fig:comp_time}
\end{figure}

Finally, in Fig. \ref{fig:comp_time}, we discuss the computational time performance of the optimal and heuristic schemes to provide some insights about their feasibility in real-world scenarios. Specifically, in Fig. \ref{subfig:time_opt_heu}, we compare the scalability performance by showing the time required to compute a rescheduling solution as the number of microservices increases from $M=10$ to $M=40$. Although the scenario complexity is rather modest, the convergence time of the optimal solution rises quickly, making it a prohibitive option for large and highly dynamic microservice deployments subject to frequent re-orchestrations. In contrast, the heuristic solution proves to be a viable alternative, as it requires significantly less time and scales effectively with the growing number of microservices. Similarly, in Fig. \ref{subfig:time_heu_k8s}, we compare the resolution time as the number of cloud regions increases. Higher number of cloud regions renders the QoS fulfillment more challenging due to the higher number of co-location requirements, prolonging the resolution time. Overall, this configuration parameter has a moderate impact on the scalability performance of both schemes. Nonetheless, the heuristic solution is way less affected than the optimal scheme, which still suffers from a visible performance degradation. This result showcases the computational efficiency of the proposed heuristic approach and promotes its applicability on practical multi-cloud environments.

\begin{table}[t] 
\centering
\caption{Control plane latency}
\begin{tabular}{| c | c | c |}
\hline
\textbf{Software framework} & \textbf{2 regions} & \textbf{5 regions}\\ 
\hline
K8s + custom plugin & 59ms & 78ms\\ 
Default K8s & 50ms & 51ms\\
\hline
\end{tabular}
\label{tab2:control_plane}
\end{table}

\subsection{Feasibility and limitations in real-world scenarios}

Beyond the performance improvements provided by the proposed algorithmic approach, the architectural design of the rescheduling scheme ensures a transparent mechanism for re-orchestrating microservices without interfering with Kubernetes' built-in rescheduling functionalities. 
Specifically, events interrupting service availability such as microservice crashes, node failures, network communications errors are automatically handled by Kubernetes control plane, which evicts and redeploys unresponsive microservices. In this regard, our solution adopts a similar reactive approach that can be triggered on-demand by the service provider once Kubernetes' rescheduling mechanism has completed. This ensures that the rescheduling policy is computed based on a stable microservice deployment, thus avoiding operational conflicts with Kubernetes workflow. Furthermore, by integrating our design as a Kubernetes scheduler plugin, we provide a seamless rescheduling process with minimal control plane overhead. In this regard, as shown in Table \ref{tab2:control_plane}, we report the average time required to compute the  rescheduling of one microservice (i.e., steps 2–4 in Fig. \ref{fig:architecture}) with $R=2$ and $R=5$ cloud regions when the number of deployed microservices is $N=400$. The measurements indicate a minimal latency overhead compared to the default Kubernetes control plane. However, this additional delay is practically insignificant when compared to the rescheduling time (i.e., step 5 in Fig. \ref{fig:architecture}), as shown by the results. Therefore, it has no negative impact on overall performance. The developed plugin leverages customized Kubernetes APIs to retrieve microservice deployments and nodes information, making informed rescheduling decisions without introducing additional complexity. This aspect combined with the scaling efficiency of the proposed rescheduling algorithm enables its feasibility in large-scale deployments with many nodes.

However, while our approach offers several advantages, we acknowledge two main limitations that impact performance in practical deployments. The proposed rescheduling procedure assumes that the microservice reboot time is homogeneous. However, its duration can vary based on multiple factors, including container image size, network configurations, and microservice dependencies.  
Longer reboot times can degrade QoS by creating synchronization issues among rescheduled microservices. Although we mitigate this downside by minimizing rescheduling duration, this solution may be less effective when microservices reboot times become increasingly heterogeneous. In addition to this, our scheme ignores the challenges related to rescheduling stateful microservices. Specifically, state management techniques such as state check-pointing, data replication, and transactional rollback mechanisms are required  to coherently and safely migrate microservices' status between nodes. For this reason, the proposed solution should be restricted in rescheduling only stateless microservices that do not share dependencies with stateful ones in order to preserve data consistency and transactional integrity of the applications. Future work will focus on addressing these limitations, particularly enhancing state management capabilities to support the rescheduling of stateful microservices more effectively.

\section{Conclusion}
We considered the problem of cost minimization in multi-cloud environments, that consists of differently-priced and geographically distributed virtual nodes offered by different cloud providers.  In this context, the variable resource requirements and life-cycle dynamism typical of microservice deployments progressively deteriorate the resource utilization efficiency, leading to higher deployment costs. To address this issue, we designed a re-orchestration scheme that reschedules deployed microservices in order to reduce the resource fragmentation, thereby reducing the deployment cost. At the same time, the scheme mitigates service disruptions caused by the rescheduling process, which temporarily affect applications' dependability, and preserves microservice colocation requirements within the same cloud region. In particular, we enhanced service continuity by combining the use of a rolling-update deployment strategy with the minimization of the number of rescheduled microservices and rescheduling duration. We analytically incorporated the service disruption model within the deployment cost model  and we formulated  a multi-objective linear integer problem to compute the optimal rescheduling solution satisfying the aforementioned objectives.
We approximated the optimal solution computation by designing a heuristic scheme based on a greedy iterative approach that ensures a balanced trade-off between cost minimization and service disruption. We evaluated the performance of the optimal and heuristic schemes by integrating both approaches as a custom Kubernetes scheduler plugin. 
The results show that our approach provides the best compromise between cost minimization, service disruption mitigation, and QoS preservation  compared to the benchmark schemes.

\section*{Acknowledgements}
This work has received funding from the EU Horizon Europe R\&I Programme under Grant Agreement no. 101070473 (FLUIDOS).

\bibliographystyle{IEEEtran}
\bibliography{bibliography}

\vskip -2.5\baselineskip plus -0.95fil
\begin{IEEEbiography}[{\includegraphics[width=0.95in,height=1.25in,clip,keepaspectratio]{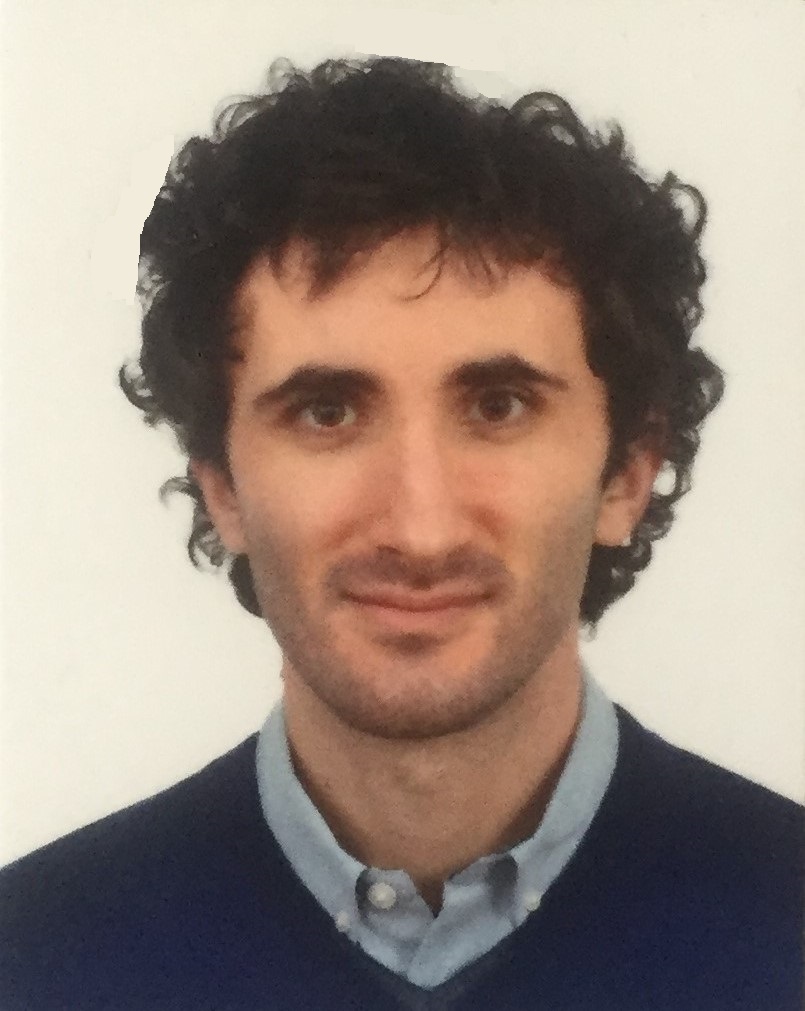}}] {Marco Zambianco}
 is a researcher at Fondazione Bruno Kessler (FBK) in Trento, Italy.
He received his Master’s Degree in Telecommunication Engineering from University of Padova and his Ph.D. (with honors) in Information Technology from Politecnico di Milano in 2022. His research interests include microservice orchestration in distributed cloud computing and cybersecurity. 
\end{IEEEbiography}
\vskip -2.5\baselineskip plus -0.95fil
\begin{IEEEbiography}[{\includegraphics[width=1in,height=1.3in,clip,keepaspectratio]{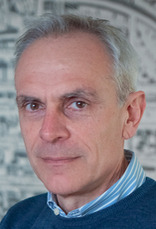}}] {Silvio Cretti} is a software architect at Fondazione Bruno Kessler (FBK) in Trento, Italy. He received his Master’s Degree in Physics at University of Trento. Before joining FBK, he worked as software engineer, architect and technical leader for Telecom Italia group. His main technical interests are related to distributed and heterogeneous clouds, container orchestration, workload placement and cloud-native applications.
\end{IEEEbiography}
\vskip -2.5\baselineskip plus -0.95fil
\begin{IEEEbiography}[{\includegraphics[width=1in,height=1.25in,clip,keepaspectratio]{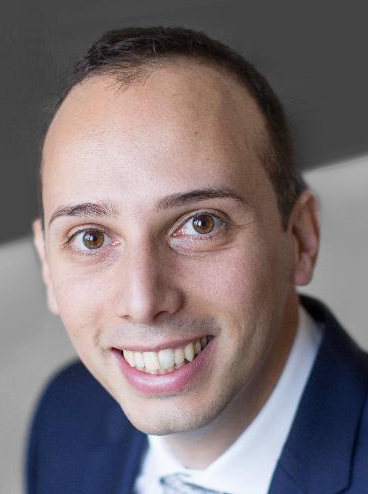}}]{Domenico Siracusa}
is associate professor at the University of Trento. Previously, he was the head of the Robust and Secure Distributed Computing (RiSING) research unit at Fondazione Bruno Kessler (FBK). He received his PhD in Information Technology at Politecnico di Milano in 2012 and earned an Executive MBA in Business Innovation at the MIB Trieste School of Management in 2022. His research interests include infrastructure security and robustness, service orchestration and management, cloud and fog computing, and SDN/NFV and virtualization. Domenico authored more than 100 publications appeared in international peer reviewed journals and in major conferences on computing and networking technologies. Domenico was project manager and technical leader of relevant EU-funded projects, such as the H2020 EU-Korea DECENTER, H2020 ACINO and EIT Digital DigiFlow projects, and was involved as principal investigator, workpackage leader or contributor in other FP7/H2020, EIT digital and commercial projects.
\end{IEEEbiography}

\end{document}